\documentclass[twocolumn]{aastex63}
\usepackage{multirow,amsmath,amsthm,mathtools}
\usepackage{chngcntr,amssymb}
\usepackage{natbib}
\usepackage{color}
\usepackage{hyperref} 
\usepackage{lineno}
\usepackage[utf8]{inputenc}
\usepackage[T1]{fontenc}

\def	\cm		{\,{\rm {cm}}}
\def	\K		{\,{\rm K}}
\def	\g		{\,{\rm {g}}}
\def	\mum	{\,{\mu \rm{m}}}
\def \Td {T \rm{_d}}

\def \NHt {N(\rm H_2)}
\def \S {\mathcal{S}}
\def \Ma {\mathcal{M}_A}
\def \Sun {\odot}
\def \aalign {a_{\rm align}}

\def \bea {\begin{eqnarray}}
\def \ena {\end{eqnarray}}

\def	\B	{{\rm B}}

\def	\C	{{\rm C}}

\def	\cm	{\,{\rm cm}}
\def	\d	{{\rm d}}

\def	\D	{{\rm D}}

\def	\g	{\,{\rm g}}

\def	\G	{{\rm G}}

\def	\H	{{\rm H}}

\def	\N	{{\rm N}}
\def	\nH	{n_{\rm H}}

\def	\pc	{\,{\rm pc}}

\def	\s	{\,{\rm s}}

\def	\V	{{\rm V}}

\accepted{2023}

\shorttitle{BALLAD-POL: I. Magnetic fields and dust in G11.11-0.12}

\graphicspath{{./}{figures/}}

\begin{document}

\title{\LARGE{\textbf{B-fields And dust in interstelLar fiLAments using Dust POLarization (BALLAD-POL): I. The massive filament G11.11$-$0.12 observed by SOFIA/HAWC+}}}

\author[0000-0002-5913-5554]{Nguyen Bich Ngoc}
\affiliation{Department of Astrophysics, Vietnam National Space Center, Vietnam Academy of Science and Technology, 18 Hoang Quoc Viet, Hanoi, Vietnam}
\affiliation{Graduate University of Science and Technology, Vietnam Academy of Science and Technology, 18 Hoang Quoc Viet, Hanoi, Vietnam}

\author[0000-0002-2808-0888]{Pham Ngoc Diep}
\affiliation{Department of Astrophysics, Vietnam National Space Center, Vietnam Academy of Science and Technology, 18 Hoang Quoc Viet, Hanoi, Vietnam}
\affiliation{Graduate University of Science and Technology, Vietnam Academy of Science and Technology, 18 Hoang Quoc Viet, Hanoi, Vietnam}

\author[0000-0003-2017-0982]{Thiem Hoang}
\affiliation{Korea Astronomy and Space Science Institute, 776 Daedeokdae-ro, Yuseong-gu, Daejeon 34055, Republic of Korea}
\affiliation{University of Science and Technology, Korea, 217 Gajeong-ro, Yuseong-gu, Daejeon 34113, Republic of Korea}

\author[0000-0002-6488-8227]{Le Ngoc Tram}
\affiliation{Max-Planck-Institut f\"{u}r Radioastronomie, Auf dem H\"{u}gel 69, 53-121, Bonn, Germany}

\author{Nguyen Chau Giang}
\affiliation{Korea Astronomy and Space Science Institute, 776 Daedeokdae-ro, Yuseong-gu, Daejeon 34055, Republic of Korea}
\affiliation{University of Science and Technology, Korea, 217 Gajeong-ro, Yuseong-gu, Daejeon 34113, Republic of Korea}

\author[0000-0003-1990-1717]{Ng\^an L\^e}
\affiliation{Institute of Astronomy, Faculty of Physics, Astronomy and Informatics, Nicolaus Copernicus University, Grudzi\k{a}dzka 5, 87-100 Toru\'{n}, Poland}

\author[0000-0002-3437-5228]{Thuong Duc Hoang}
\affiliation{Kavli Institute for the Physics and Mathematics of the Universe (Kavli IPMU, WPI), UTIAS, The University of Tokyo, Kashiwa, Chiba 277-8583, Japan}

\author[0000-0002-4372-5509]{Nguyen Thi Phuong}
\affiliation{Korea Astronomy and Space Science Institute, 776 Daedeokdae-ro, Yuseong-gu, Daejeon 34055, Republic of Korea}
\affiliation{Department of Astrophysics, Vietnam National Space Center, Vietnam Academy of Science and Technology, 18 Hoang Quoc Viet, Hanoi, Vietnam}

\author[0000-0001-9657-8728]{Nguyen Minh Khang}
\affiliation{Institute For Interdisciplinary Research in Science and Education (IFIRSE), ICISE, 07 Science Avenue, Ghenh Rang Ward, 55121 Quy Nhon City, Binh Dinh Province, Vietnam}

\author[0000-0002-5678-1008]{Dieu D. Nguyen}
\affiliation{Université de Lyon1, Ens de Lyon, CNRS, Centre de Recherche Astrophysique de Lyon (CRAL) UMR5574, F-69230 SaintGenis-Laval, France}

\author[0000-0001-9654-8051]{Gia Bao Truong Le}
\affiliation{Korea Astronomy and Space Science Institute, 776 Daedeokdae-ro, Yuseong-gu, Daejeon 34055, Republic of Korea}
\affiliation{University of Science and Technology, Korea, 217 Gajeong-ro, Yuseong-gu, Daejeon 34113, Republic of Korea}

\correspondingauthor{Pham Ngoc Diep} \email{pndiep@vnsc.org.vn}

\begin{abstract}
We report the first measurement of polarized thermal dust emission toward the entire Infrared Dark Cloud G11.11$-$0.12 taken by the polarimeter SOFIA/HAWC+ at 214 $\mum$. The obtained magnetic fields (B-fields) from the polarized emission of the early-stage and massive filament tend to be perpendicular to its spine. We produce a map of B-field strengths for the center region of the filament. The strengths vary in the range of 100-600 $\mu\rm{G}$ and are strongest along the filament's spine. The central region is sub-Alfv\'enic and mostly sub-critical, meaning that B-fields dominate over turbulence and are strong enough to resist gravitational collapse. The alignment and properties of dust grains in the filament are studied using RAdiative Torque (RAT) theory. We find the decrease of polarization degree $P$ with emission intensity $I$, i.e., depolarization effect, of the form $P\propto I^{-\alpha}$ with $\alpha\sim$0.8-0.9, implying a significant loss of grain alignment in the filament's spine. The depolarization can be explained by the decrease in RAT alignment efficiency toward the denser regions with weaker radiation field, which cannot be explained by B-field tangling. We study the effect of the enhanced magnetic relaxation by embedded iron inclusions on RAT alignment and find that the high polarization fraction $P\sim$20-30\% in the outer layer of the filament is potential evidence for the magnetically enhanced RAT alignment mechanism. This is the first time this effect is evaluated in a filament. Based on the polarization fraction and RAT alignment theory, we also find evidence for grain growth in the filament.
\end{abstract}


\section{Introduction}\label{sec:intro}
Observations by {\it Herschel} reported the ubiquity of filaments in the ISM where prestellar cores and protostars are formed \citep{andre2014filamentary}. It suggests that filaments can be important stages in the star-forming process. In their earliest phases, several cores can be observed along filaments. Infrared-dark clouds (IRDCs) are cradles for the formation of stars, especially high-mass stars and star clusters(e.g., \citealt{rathborne2006IRDCs}). IRDCs are quiescent molecular clouds containing mostly cold and dense molecular gas. The IRDCs would be fragmented to form a number of compact cores and clumps. As they evolve over time, bubbles or \ion{H}{2} regions form due to stellar feedback from massive stars. Then, the filaments will contain bright clusters and appear similar to the filamentary molecular clouds associated with high-mass star-forming regions. 

The role of B-fields in the formation and evolution of filaments is likely vital because it makes the flows more coherent, allowing filaments to survive longer (see \citealt{hennebelle_inutsuka2019} for a review). However, the behavior of B-fields in filaments is less well-understood, in particular, for high-mass stars in their earliest phases. B-fields, along with gravity, turbulence, and stellar feedback are believed to play essential roles in the formation of massive stars \citep{crutcher2012}.

The measurement of polarized dust emission has become an essential tool for probing B-field properties, as the dust grains are preferentially aligned with their major axes perpendicular to the local B-field direction (see \citealt{lazarian2007, andersson2015} for reviews). From the observed polarized dust emission, the B-field morphology can be inferred by rotating the dust polarization angles by 90$^\circ$. Polarimetric observations at far-infrared (far-IR) and sub-millimeter (sub-mm) wavelengths are widely used to infer the morphology and strengths of B-fields (see \citealt{pattle2019} for a review). The strengths of the plane-of-sky component of the B-fields can be commonly estimated using the Davis-Chandrasekhar-Fermi method (DCF method; \citealt{davis1951strength, chandrasekhar1953}).

Previous studies of B-fields toward interstellar filaments using polarized dust emission usually reveal that the B-fields are parallel to the filament in diffuse regions and become perpendicular to the filament's spine in dense regions when the hydrogen column density exceeds $N_{\rm H}\sim$(3-5)$\times 10^{21}\cm^{-2}$ (e.g., see \citealt{Planck.2016,Soler.2017,Soler.2019jur} for observations and \citealt{Soler.2017vv} for simulations). To date, there are only a handful of studies about B-fields in massive filaments, including G35.39$-$0.33 \citep{liu2018G35}, G34.43+0.24 \citep{tang2019gravity,soam2019G34}, G14.225$-$0.506 \citep{anez2020role},  NGC 6334 \citep{arzoumanian2021NGC6334}, and G47.06+0.26 \citep{stephens2022G47}.

This paper first aims to characterize the properties of B-fields in a high-mass star-forming filament G11.11$-$0.12 (hereafter G11 or colloquially known as the Snake filament). We use data taken by the High-resolution Airborne Wideband Camera Plus (HAWC+; \citealt{harper2018hawc}) accommodated on the Stratospheric Observatory for Infrared Astronomy (SOFIA; \citealt{temi2018sofia}) at 214 $\mum$ wavelength. We note that B-fields in the center region of G11 were studied using the JCMT/SCUPOL polarimetric observations at 850 $\mum$ and $20\arcsec$ ($0.35\;\pc$) resolution \citep{pillai2015magnetic}. They found that the B-fields toward this region are perpendicular to the filament's spine and estimated the lower limit of the plane-of-sky B-field strength to be $\sim200\;\mu\rm{G}$. This value is converted from the total B-field strength using an average field geometry with a conversion factor of 1.3 \citep{crutcher2004}.

G11 is one of the IRDC filaments in the galactic plane on the near side of the Scutum-Centaurus arm \citep{wang2014hierarchical}. It is cold, dense, and in an early phase of evolution \citep{jackson2010nessie} in contrast with its late-phase counterpart such as Orion or NGC 6334. G11 is located at a distance of \mbox{3.6 kpc} from Earth \citep{pillai2006a}, $22\;\pc$ long, 1.0 pc wide, and has a mass of $1.5 \times 10^{4}\;M_\Sun$ \citep{zucker2018filament}. The filament has a linear mass density of $\sim$$600~M_\Sun/\pc$, already fragmented, and new stars are expected to be in the process of forming in the filament \citep{kainulainen2013high}. G11 is identified to host seven JCMT/SCUBA clumps \citep{johnstone2003g11} and 18 protostellar cores along its spine \citep{henning2010seeds}. There are two massive clumps in the filament (P1 and P6 in Figure \ref{fig:Bfieldlength}) with a mass of $\sim$$1000\;M_\Sun$ and sizes less than $1~\pc$, which are the sites for high-mass star formation \citep{wang2014hierarchical}. All these above reasons make G11 an ideal target for studying the initial conditions for the formation of IRDCs with the existence of B-fields and, in a broader scope, B-fields in high-mass star-forming regions in their earliest phases.

While dust polarization angles are useful for measuring B-field morphology and strengths, dust polarization fraction provides crucial constraints on dust physics (including grain alignment and disruption) and grain properties (size, shape, composition). The leading theory of grain alignment is based on radiative torques (RATs, \citealt{lazarianhoang2007,Hoang.2016, hoang2022internal}). The presence of iron inclusions embedded in dust grains is predicted to enhance magnetic relaxation and increases the RAT alignment degree, aka magnetically enhanced RAT alignment or MRAT \citep{Hoang.2016}. The RAT theory was observationally tested in different environments, using starlight polarization (see \citealt{andersson2015} for a review) and polarized thermal dust emission in star-forming regions \citep{tram2020understanding,tram2022recent}. Moreover, the RAT theory has been widely used to interpret observations such as \cite{planck2018} and \cite{ward2017first}. The massive G11 filament is an ideal target for testing the RAT theory due to the lack of bright embedded sources in its early phase, therefore, the only significant radiation source impacting the filament is the interstellar radiation field. In addition to that, dust polarization in the dense region can allow us to constrain grain growth \citep{Vaillancourt:2020ch,hoang2021polhole}. Thus, the second aim of our paper is to probe the properties of dust grains in G11. We perform a comprehensive analysis of dust polarization fraction to constrain dust physics and dust properties. In particular, with the measured map of B-field strengths, it is possible to study the efficiency of magnetic relaxation and the role of the MRAT mechanism.

This paper is the first one in our series aiming at characterizing the properties of B-fields And dust in interstelLar fiLAments using Dust POLarization (BALLAD-POL).

The structure of the paper is as follows. In Section \ref{sec:obs}, we describe the observations and analyses of thermal dust polarization emission. Section \ref{sec:Bfield} is devoted to the study of B-field morphology and strengths. In Section \ref{sec:dust}, we study the grain alignment physics and dust properties using the polarization fraction. Our results are discussed in Section \ref{sec:dis}. Finally, our conclusions are presented in Section \ref{sec:conclusions}.

\section{Observations} \label{sec:obs}

In this work, we use the archival far-IR polarimetric data observed by SOFIA/HAWC+ at 214$\mum$ wavelength toward G11 as part of the FIELDMAPS legacy project (PI: Ian W. Stephens, Worcester State University). The angular resolution is $18.\arcsec2$ corresponding to a physical scale of 0.32 $\pc$ at a distance of $\sim$\mbox{3.6 kpc}. G11 was mapped by mosaicking four fields of SOFIA/HAWC+. The observations were taken on 13 July 2018 with a total exposure time of $2316 \s$. Polarimetric observations were performed with the chop-nod imaging mode \citep{harper2018hawc}. We use the Level 4 data products accessible from the SOFIA archive\footnote{https://irsa.ipac.caltech.edu/Missions/sofia.html}. The data reduction process was carried out using \href{https://www.sofia.usra.edu/sites/default/files/2022-12/hawc_users_revL.pdf}{HAWC DRP PIPELINE v1.3.0} in several steps including chop-nop subtraction, flat-field correction, flux calibration, background subtraction, polarization angle correction, and mosaic map making.

For linear polarization, the polarization states are defined by the Stokes $I$, $Q$, and $U$. The original means of signal-to-noise ratios ($S/N$s) of $I$, $Q$, and $U$ are 28.3, 0.2, and 0.5, respectively. Due to the low $S/N$s of the data, we binned the original $4\arcsec.55$-pixels to $9\arcsec.1$ to increase the $S/N$s. The new pixel-size is about a half of the beam size of SOFIA/HAWC+ at 214$\mum$. After binning, the mean $S/N$s increase by a factor of two: 57.4 for $I$, 0.4 for $Q$, and 1.0 for $U$, respectively. Rigorous cuts will be applied to the data for further analyses which are mentioned elsewhere in the paper.

\begin{figure*}[!htb]
\centering
\includegraphics[trim=0cm 0cm 0cm 0cm,clip,width=11.5cm]{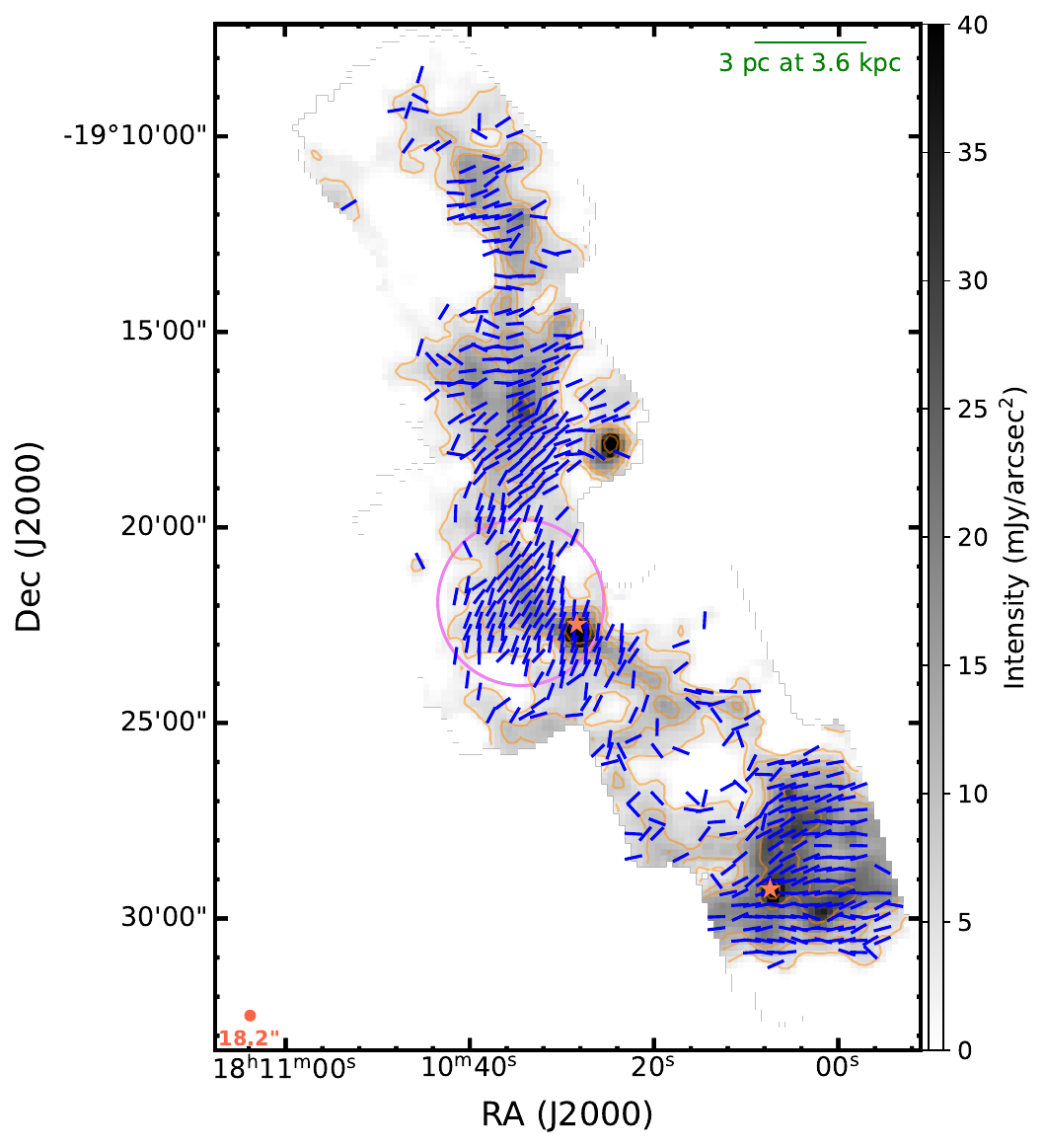}
\caption{Map of B-field orientation toward G11 at $214 \mum$ measured by SOFIA/HAWC+. The blue segments display the B-field orientation. The B-field segments are plotted such that the spacing between neighboring ones is equal to the beam size. The beam size of SOFIA/HAWC+ at $214 \mum$ is $18\arcsec.2$, which is shown by the red circle in the lower-left corner. The B-fields are over-plotted on a gray map of intensities ($I$). The orange contours are for $I= 3.6,8.4,12.0,24.0,42.0$ mJy/arcsec$^2$. The red stars mark the locations of two massive protostellar candidates: P1 in the central and P6 in the southern regions \citep{henning2010seeds}. The violet circle shows the region observed by JCMT/SCUPOL whose B-fields are shown in Figure \ref{fig:bfieldscupol}.} \label{fig:Bfieldlength} 
\end{figure*}

The observed intensity for each pixel is calculated by
\begin{eqnarray}
PI_{\rm obs} = \sqrt{Q^2 + U^2}.
\end{eqnarray} 

Because of the presence of the noise in $Q$ and $U$ in $PI_{\rm obs}$, we debias $PI_{\rm obs}$ by using the following relation \citep{serkovski1974}

\begin{eqnarray}
PI=\sqrt{Q^2+U^2-0.5(\delta Q^2 + \delta U^2)},
\end{eqnarray} 
where the uncertainty on $PI$ is
\begin{eqnarray}
\delta PI=\sqrt{\frac{Q^2\delta Q^2+U^2\delta U^2}{Q^2+U^2}}.
\end{eqnarray}

The polarization fraction, $P$, and its uncertainty, $\delta P$, are calculated as 
\begin{eqnarray}
P(\%)=100\times\frac{PI}{I},
\end{eqnarray}
and
\begin{eqnarray}
\delta P(\%)=100\times \sqrt{\frac{\delta PI^2}{I^2}+\frac{\delta I^2(Q^2+U^2)}{I^4}}. 
\end{eqnarray}

We calculate the polarization angle, $\theta$, and its uncertainty, $\delta \theta$, by using:
\begin{eqnarray}
\theta=0.5\tan^{-1}\left(\frac{U}{Q}\right),
\end{eqnarray}
and
\begin{eqnarray}
\delta \theta=0.5\times\frac{\sqrt{U^2\delta Q^2+Q^2\delta U^2}}{(Q^2+U^2)}.
\end{eqnarray}

The B-field angles on the plane of the sky (POS) are obtained by rotating the polarization angles by $90^\circ$. Following the IAU convention, the angle of B-fields is east of north, ranging from 0$^\circ$ to 180$^\circ$.

For further analyses, we only use the data satisfying: $S/N(I)>10$, $S/N(P)>3$, $S/N(\theta)>3$, and \mbox{$P<30\%$}. The resulting B-field orientation map is displayed in Figure \ref{fig:Bfieldlength}.
 
\begin{figure}[!htb]
\centering
\includegraphics[width=8.5cm]{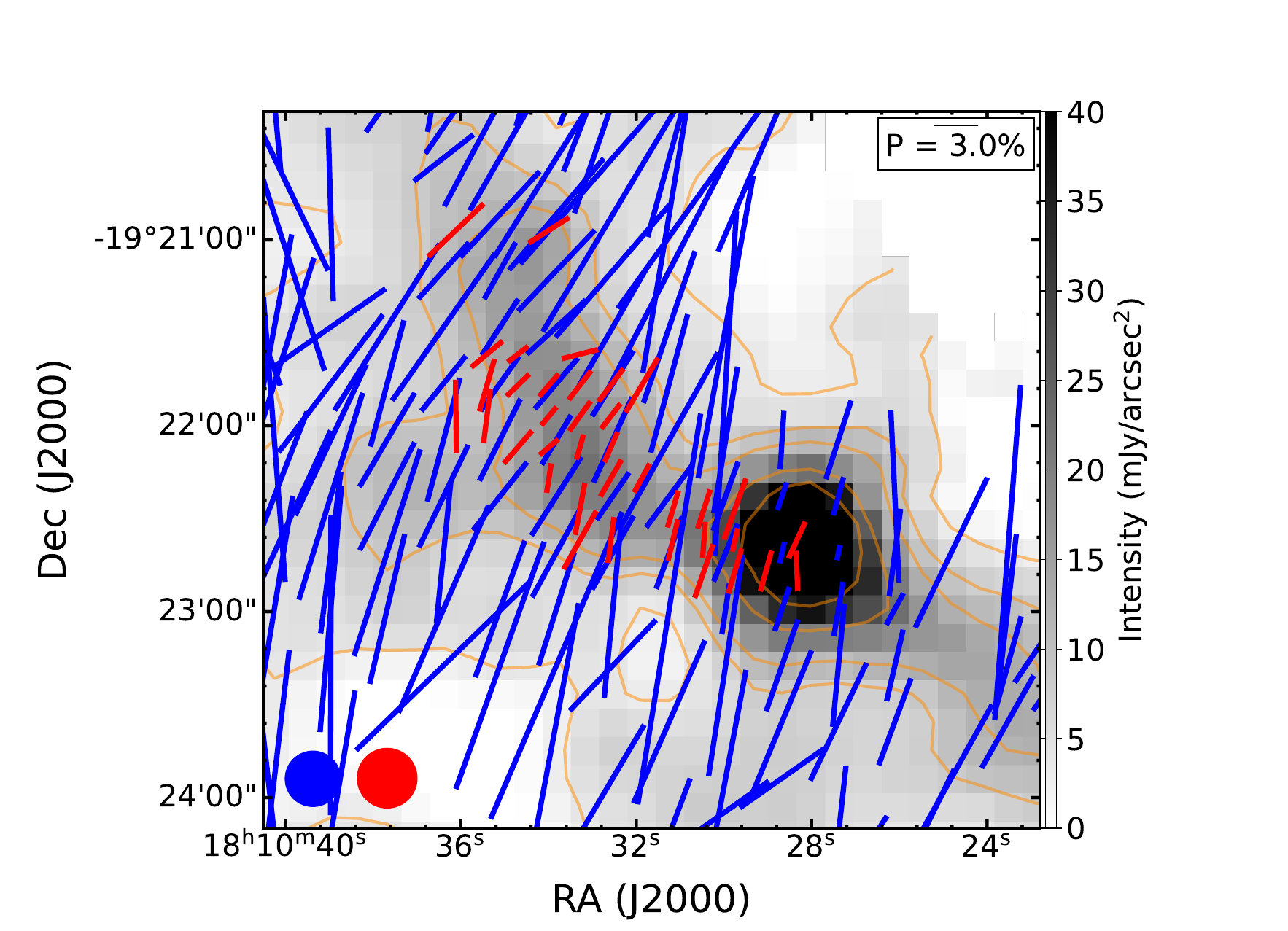}
\caption{Comparison of the B-field orientation obtained by SOFIA/HAWC+ at 214 $\mum$ (blue segments) and JCMT/SCUPOL (red segments) at 850 $\mum$. The lengths of B-field segments are proportional to the polarization fraction. A reference segment of 3\% is given in the upper-right corner. The $18\arcsec.2$ beam size of SOFIA/HAWC+ (blue circle) and $20\arcsec$ of JCMT/SCUPOL (red circle) are shown in the lower-left corner. The contours are the same as in Figure \ref{fig:Bfieldlength}.} \label{fig:bfieldscupol} 
\end{figure}

To compare the B-field map obtained by SOFIA/HAWC+ with the one by JCMT/SCUPOL\footnote{https://www.cadc-ccda.hia-iha.nrc-cnrc.gc.ca/en/community/scupollegacy/} \citep{matthews2009SCUPOL, pillai2015magnetic}, we plot in Figure \ref{fig:bfieldscupol} the B-field segments of the two observations together toward the center region where the observations with JCMT/SCUPOL are available. The cuts applied on the JCMT/SCUPOL data are $S/N(P)>3$, $\delta \theta< 8^\circ$, and $P<5\%$. There is a general agreement on the B-field orientation of two observations. However, the polarization fraction obtained by SOFIA/HAWC+ is significantly greater than that of JCMT/SCUPOL. The reasons for the difference could be due to the systematic of the two facilities and the dependence of polarization fraction on temperatures, densities, and dust composition (whether separate or mixed) at different wavelengths \citep{Lee.2020,tram2020understanding}. Indeed, the shorter wavelength emission collected by SOFIA/HAWC+ at $214 \mum$ traces the warm dust and the warm dust corresponds to the higher radiation field. According to the RAT theory, the dust irradiated by higher radiation field is better aligned with B-fields, i.e. higher polarization fraction, than that of the cold dust. However, this assessment is qualitative, and finding the true answer for the difference is beyond the scope of the current studies.

\section{Magnetic Field Morphology and Strengths}\label{sec:Bfield}
\subsection{Magnetic Field Morphology}\label{subsec:Morphology}

\begin{figure}[!htb]
\centering
\includegraphics[trim=0cm 0cm 0cm 0cm,clip,width=8.5cm]{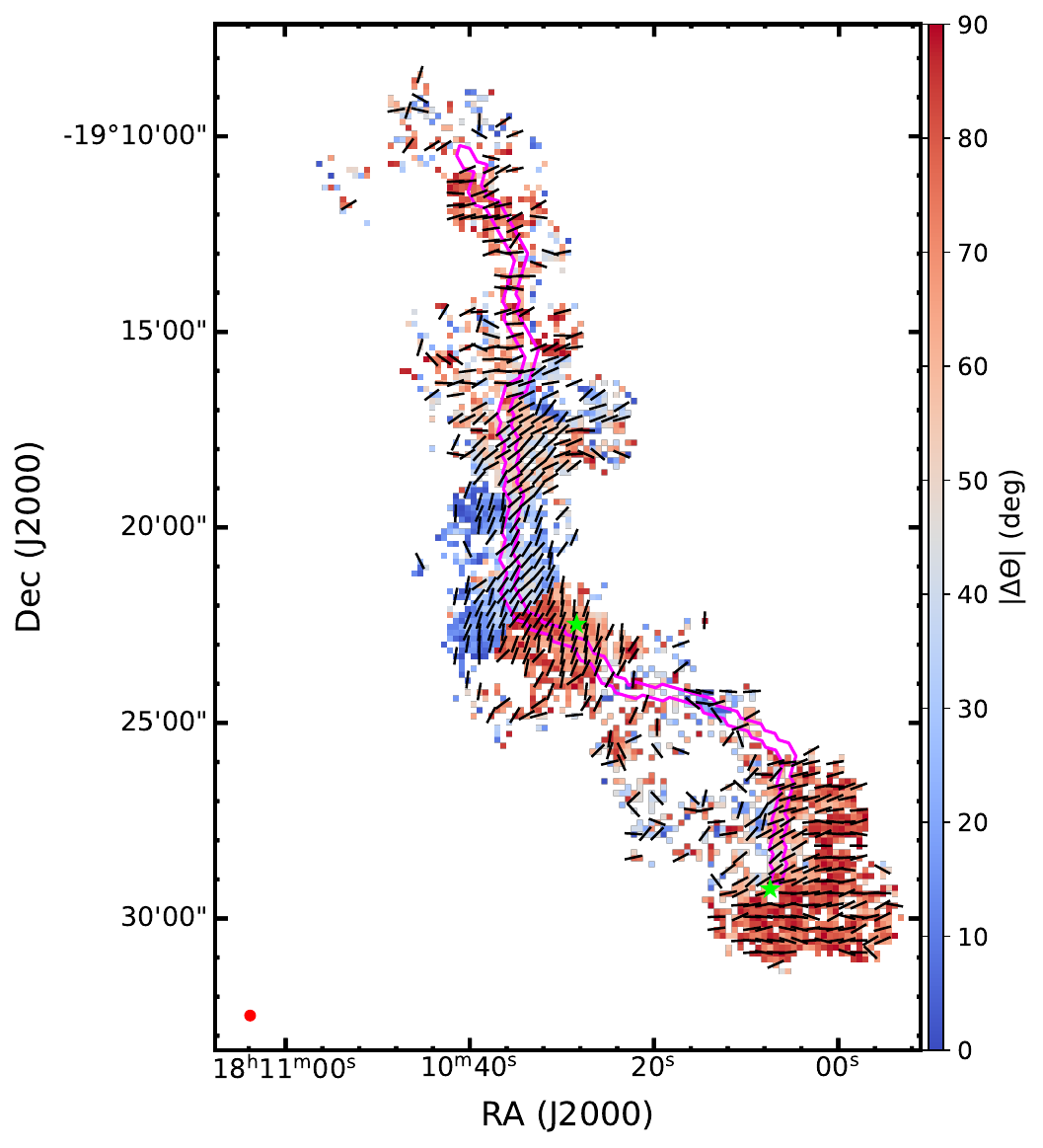}
\caption{Map of the absolute angle differences between the B-field orientation angle and the filament's spine direction (see text). The magenta contour shows the spine. The black segments display the B-field orientation with the segment length proportional to the polarization fraction of the corresponding pixel. A reference segment of 10\% is given in the upper-right corner of the figure. The green stars mark the locations of P1 and P6.}\label{fig:orien} 
\end{figure}

\begin{figure}[!htb]
\centering
\includegraphics[trim=0.7cm 0cm 0.7cm 0cm,clip,width=8.6cm]{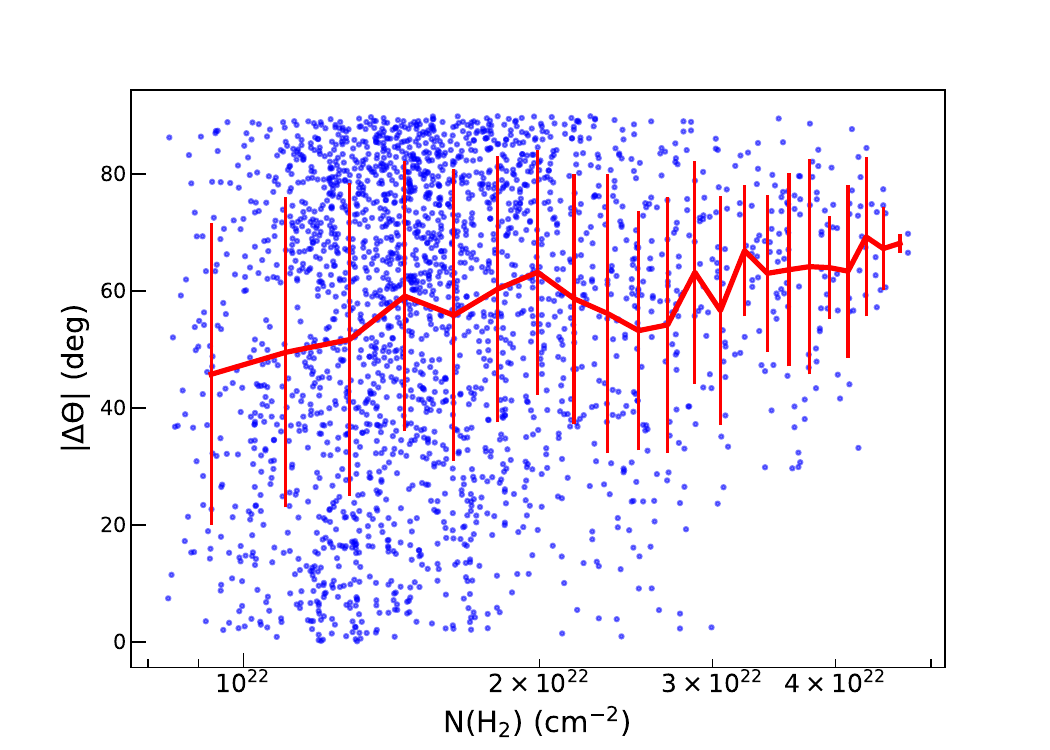}
\caption{Dependence of $|\Delta \Theta|$ on column densities. The thick red curve shows the running means of $|\Delta \Theta|$ and the error bars are their RMSs calculated for a bin size of $1.77\times10^{21}~$cm$^{-2}$.}\label{fig:orienNH2}
\end{figure}

The evolution of filament imprints on the morphology of the embedded B-fields. Magneto-hydrodynamics simulations predict that B-fields are perpendicular to the filament's spine \citep{Soler.2017vv}. We aim to quantitatively test this with G11. Firstly, we use the filament's spine identified by $RadFil$ \citep{zucker2018radfil}, the magenta contour in Figure \ref{fig:orien}. A mask of the filament’s column density map using a threshold $\approx$$2\sigma$ above the mean background is supplied to {\it RalFil}. It then topologically finds the spine using medial axis skeletonization on the mask: the set of points that has more than one closest point on the mask’s boundary (\citealp{koch2015filament} and \citealp{zucker2018radfil}). Then, for each pixel, we find its closest point to the spine and calculate the angle difference, $\Delta \Theta$, between the B-field orientation angle of the pixel and the spine angle --tangent of the spine at the closest point. Figure \ref{fig:orien} shows a map of the absolute angle differences, $|\Delta \Theta|$, indicating that the B-field orientations are nearly perpendicular to the spine for a major fraction of the total pixels (red pixels in the map). Figure \ref{fig:orienNH2} shows the dependence of $|\Delta \Theta|$ on the column densities obtained from \textit{Herschel} (see Section \ref{subsubsec:nh2}). A clear tendency of the B-fields turning from parallel to perpendicular to the filament's spine at higher column density is present, which is consistent with the findings by \citet{ade2016Bfield} and \citet{Soler.2019jur}.

\subsection{Magnetic Field Strengths and Energy Balance of the Center Region}

It was shown by \citet{davis1951strength} and \citet{chandrasekhar1953} that turbulent motions generate irregular B-fields. Based on the analysis of the small-scale randomness of B-field lines, assuming that the dispersion in the B-field angles is proportional to the Alfv\'en Mach number, the B-field strengths can be estimated, which is called the Davis-Chandrasekhar-Fermi (DCF) method. \cite{crutcher2012} proposed a variant of the method, giving an estimate of the magnitude of B-fields in the POS, $B_{\rm POS}$, as
\begin{equation}\label{equ:dcf}
B_{\rm POS}=\mathcal{Q}\sqrt{4\pi\rho}\frac{\sigma_V}{\sigma_\theta}\approx9.3\sqrt{n(\rm H_2)}\frac{\Delta V}{\sigma_\theta}\quad(\mu \rm{G}),
\end{equation}
where $\mathcal{Q} = 0.5$ is a correction factor for line-of-sight and beam-integration effects \citep{ostriker2001}, \mbox{$\rho = \mu m_H n({\rm H_2})$} is the gas density in g\;cm$^{-3}$ with mean molecular weight $\mu = 2.8$, and number densities $n(\rm H_2)$ in units of cm$^{-3}$; $\sigma_V$ the one-dimensional non-thermal velocity dispersion in km$\s^{-1}$, $\sigma_\theta$ is the dispersion of the polarization position angles about a mean B-field in degrees, $\Delta V=2.355\sigma_V$ is the full width at half maximum (FWHM) of the non-thermal velocity component in units of km$\s^{-1}$.

To obtain a map of the B-field strengths using Equation \ref{equ:dcf}, we generate three parameter maps for $\sigma_\theta$, $\Delta V$, and $n(\rm H_2)$ in the following sub-sections.

\subsubsection{Polarization Angle Dispersion}

\begin{figure}[!htb]
\centering
\includegraphics[trim=4.2cm 1.1cm 4.8cm 2.2cm,clip,width=8.5cm]{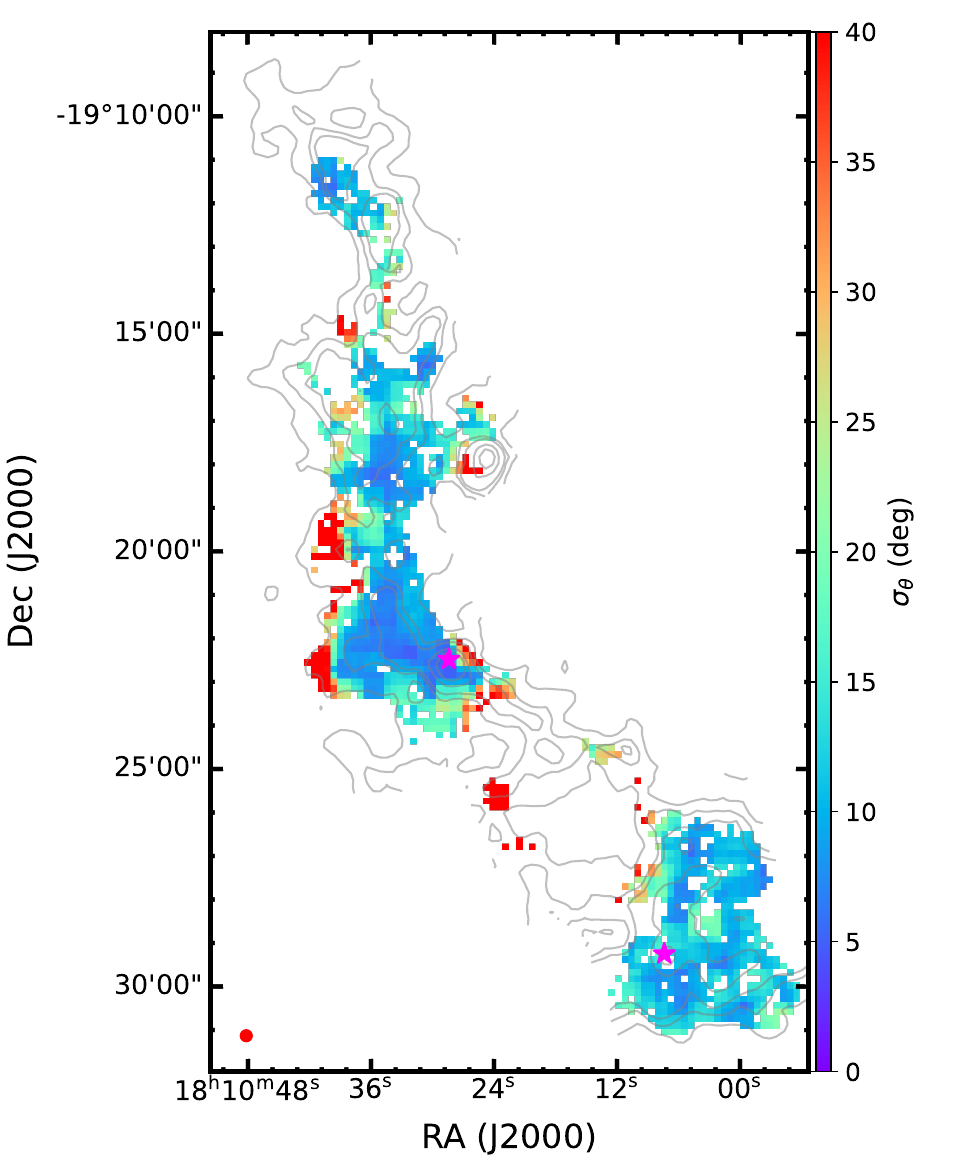}
\caption{The polarization angle dispersion map of G11. The contours are the same as in Figure \ref{fig:Bfieldlength}.} \label{fig:angdis} 
\end{figure}

To calculate the polarization angle dispersion caused by the turbulent component of the B-fields, we apply the `un-sharp masking' method which was first introduced by \citet{2017ApJ...846..122P}. Initially, the method was used to calculate the mean B-field strength of \mbox{OMC 1} of the Orion A filament. \citet{hwang2021jcmt} then used it to calculate the polarization angle dispersion for each pixel; therefore, were able to obtain a map of B-field strengths of the same region. We first calculate the mean polarization angle, $\bar{\theta}$, in a $5 \times 5$-pixel box whose size is about two beam sizes as was done in other work such as \citet{hwang2021jcmt, guerra2021orion}. In the second step, we calculate the angle dispersion, i.e. RMS of the difference between the polarization angle of pixel $i = 1, N$ of the box, $\theta_i$, and their mean value, $\bar{\theta}$, as follows:  $\sigma_\theta = \sqrt{\sum_{i=1}^N(\theta_i-\bar{\theta})^2/N}$. Here, $N$ is the included number of pixels in the box that remain from the general cut mentioned in Section \ref{sec:obs} which is required to be $N > 13$ ($>50\%$ of the total 25 pixels of the box). Then, we move the box and repeat the process over the whole map. Figure \ref{fig:angdis} shows the resulting map of the calculated polarization angle dispersion.

\subsubsection{Velocity Dispersion} \label{subsubsec:velodis}

\begin{figure}[!htb]
\centering
\includegraphics[trim=0cm 0cm 0cm 0cm,clip,width=8.5cm]{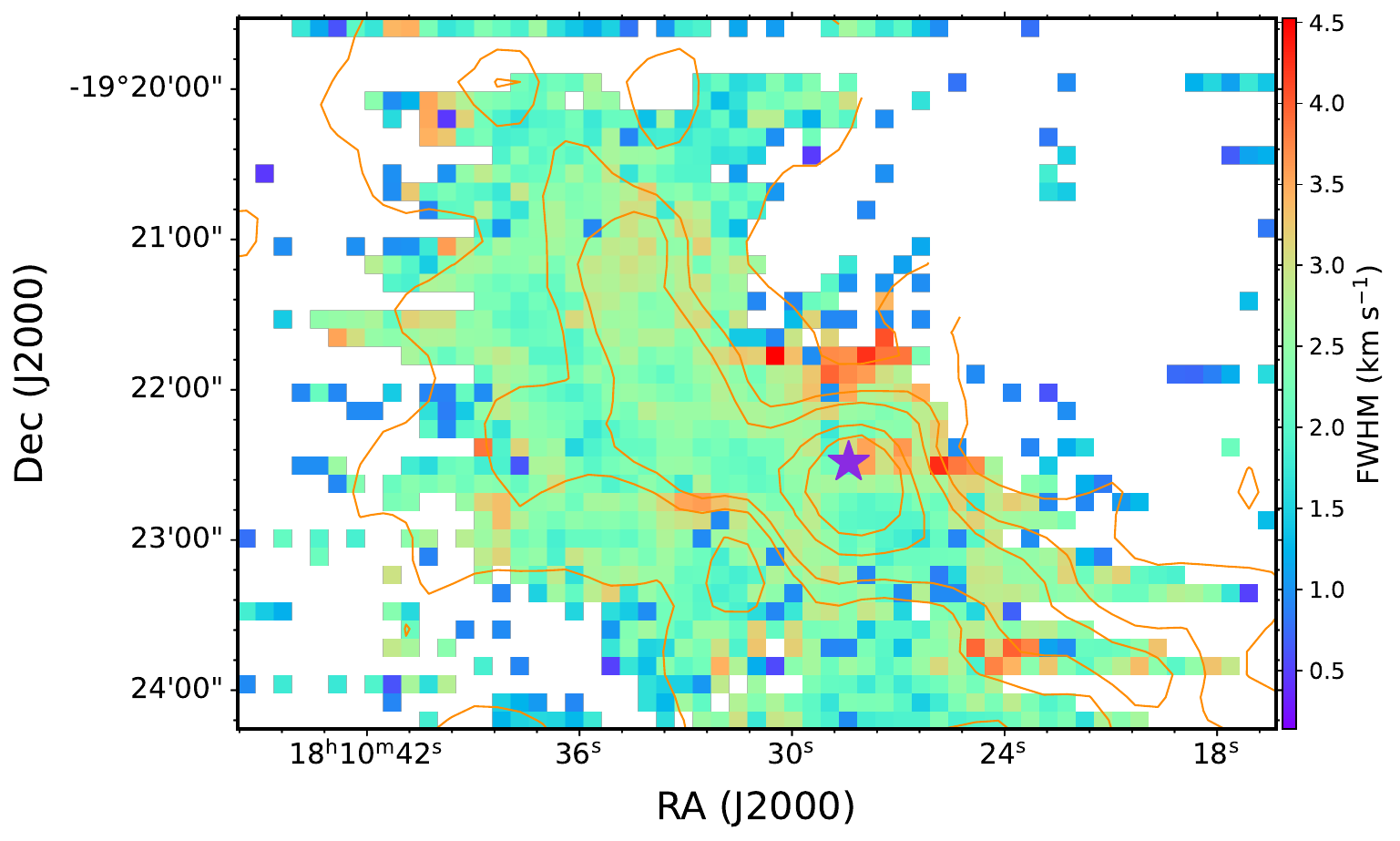}
\caption{The FWHM map of the non-thermal component obtained from $^{13}$CO spectral lines for the center region. The contours are the same as in Figure \ref{fig:Bfieldlength}. The star marker shows the location of P1.} \label{fig:fwhm} 
\end{figure}

To build a pixel-by-pixel map of the B-field strengths, we need the velocity dispersion of each pixel which requires the pixel to have high SNR and high spectral-resolution to be able to fit its spectrum with a Gaussian. Archival data of sufficient quality is only available around the vicinity of the center region.

We use $^{13}$CO($3-2$) spectral line from JCMT Science Archive\footnote{https://www.cadc-ccda.hia-iha.nrc-cnrc.gc.ca/en/jcmt/} to calculate the FWHM map of the non-thermal component, $\Delta V$. Emission from CO molecules is known to trace a wide range of H$_2$ number density between $10^2$-$10^4\cm^{-3}$ in which $^{13}$CO traces moderate density of $\sim10^3\cm^{-3}$ \citep{2019ApJ...884..137H, 2019ApJ...878..110F, minamidani2011dense}. The $^{13}$CO($3-2$) observations toward the center of G11 were carried out as part of the JCMT observations (proposal ID M15AI41). These observations were made on 2 April 2015 using HARP instrument tuned to the central frequency of 330.587 GHz with a total integration time of $1675~\s$, and atmospheric opacity ranging from 0.039 to 0.047 at 225 GHz. The data were reduced by using the ORAC Data Reduction (ORAC-DR) pipeline \citep{buckle2009} and the Kernel Application Package (KAPPA) \citep{currie2014} in Starlink. The original spectral resolution of the data set is 0.055 km$\s^{-1}$. The spectra are smoothed to have the final resolution of 0.2 km$\s^{-1}$. The spectrum of each pixel is then fitted to a Gaussian using {\it fit1d} of the SMURF package \citep{chapin2013smurf} to obtain the FWHM map of $^{13}$CO($3-2$), $\Delta V^2_{{\rm total, ^{13}CO}}$.

The non-thermal velocity component, $\Delta V$, is calculated using $\Delta V^2 = \Delta V^2_{{\rm total, ^{13}CO}} - \dfrac{k_{\rm B} T}{m_{^{13}\rm CO } }{8 \rm ln 2}$, where $m_{ ^{13}\rm CO } = 29~{\rm amu}$ is the molecular mass of $\rm ^{13}CO$, $k_{\rm B}$ is the Boltzmann constant, and $T$ is the gas temperature. The thermal contribution to the velocity dispersion is negligible if we adopt an average gas temperature of \mbox{20 K}. We show the map of the non-thermal velocity dispersion of the central region in Figure \ref{fig:fwhm}.

\subsubsection{Volume Densities} \label{subsubsec:nh2}

\begin{figure*}[!htb]
\centering
\includegraphics[trim=0cm 2cm 0cm 1cm,clip,width=17cm]{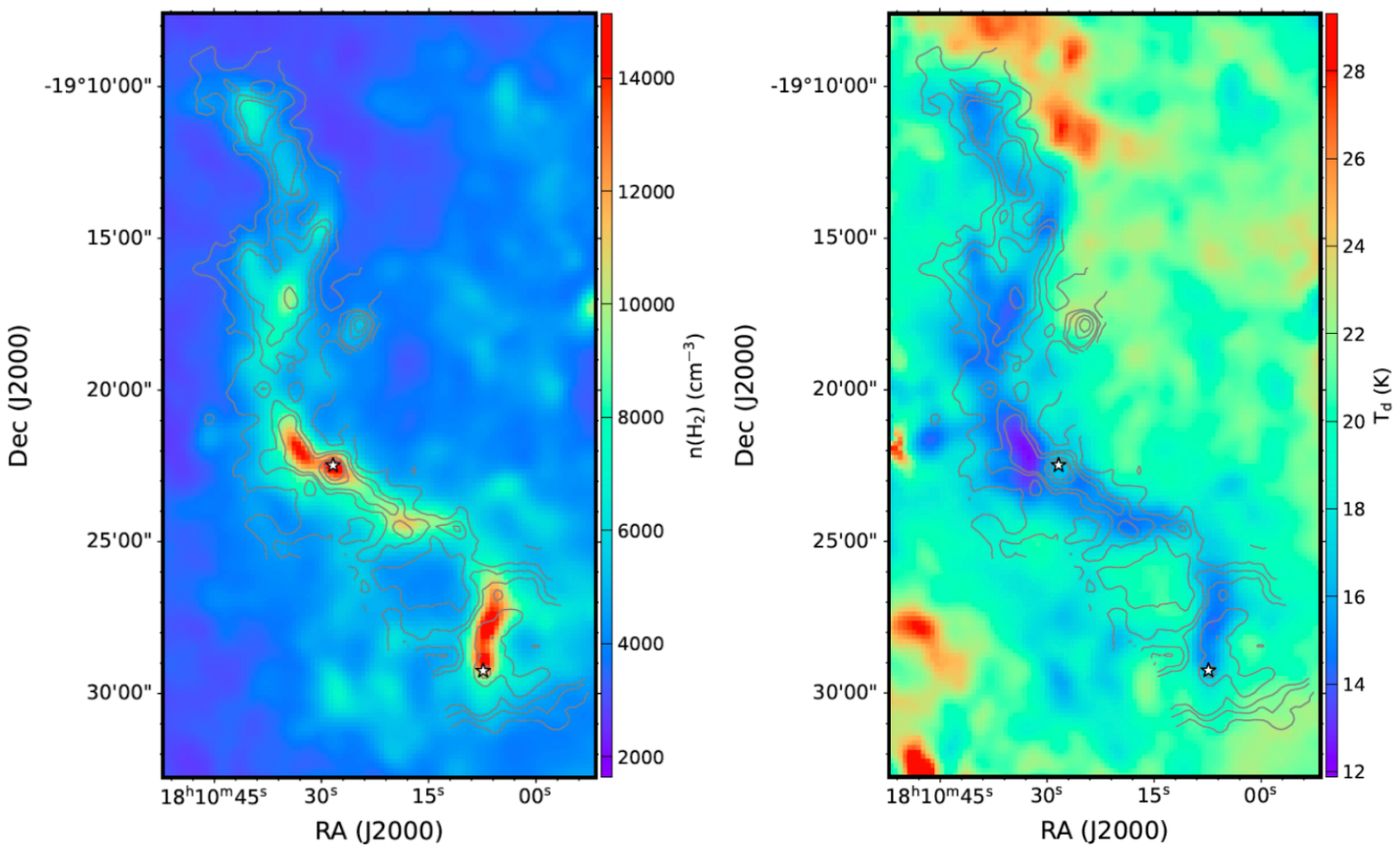}
\caption{Maps of the volume densities (left) and dust temperatures (right) of G11. Contours are the same as in Figure \ref{fig:Bfieldlength}.}. \label{fig:nH2Td} 
\end{figure*}

For calculating the volume densities, we use the maps of the column densities, $N({\rm H_2})$, and dust temperatures, $\Td$, derived by \citet{zucker2018filament} from the multi-wavelength \textit{Herschel} data. The resolution of these maps is $43\arcsec$ ($\sim$1.4$\pc$) with a pixel size of $11\arcsec.5 \times 11\arcsec.5$. \citet{zucker2018filament} also fitted the spine of G11 using {\it RadFil} algorithm \citep{zucker2018radfil} and found that filament has a length of $22 \pc$ and a width of $1.0 \pc$. We assume that the filament has a cylindrical shape so that the depth of the filament is equal to its width. The volume densities can then be calculated as follows:
\begin{equation}\label{equa:nh2}
n(\H_{2}) = \frac{N(\H_{2})}{W},
\end{equation}
where $W$ is the depth equal to $1.0 \pc$. The maps of $n(\H_{2})$ and $\Td$ are shown in Figure \ref{fig:nH2Td}. The volume densities are in the range of $10^{3}$-$10^{5}\cm^{-3}$.

\subsubsection{Magnetic Field Strengths} \label{subsec:bfieldstrength}

\begin{figure}[!htb]
\centering
\includegraphics[trim=0cm 0cm 0cm 0cm,clip,width=8.5cm]{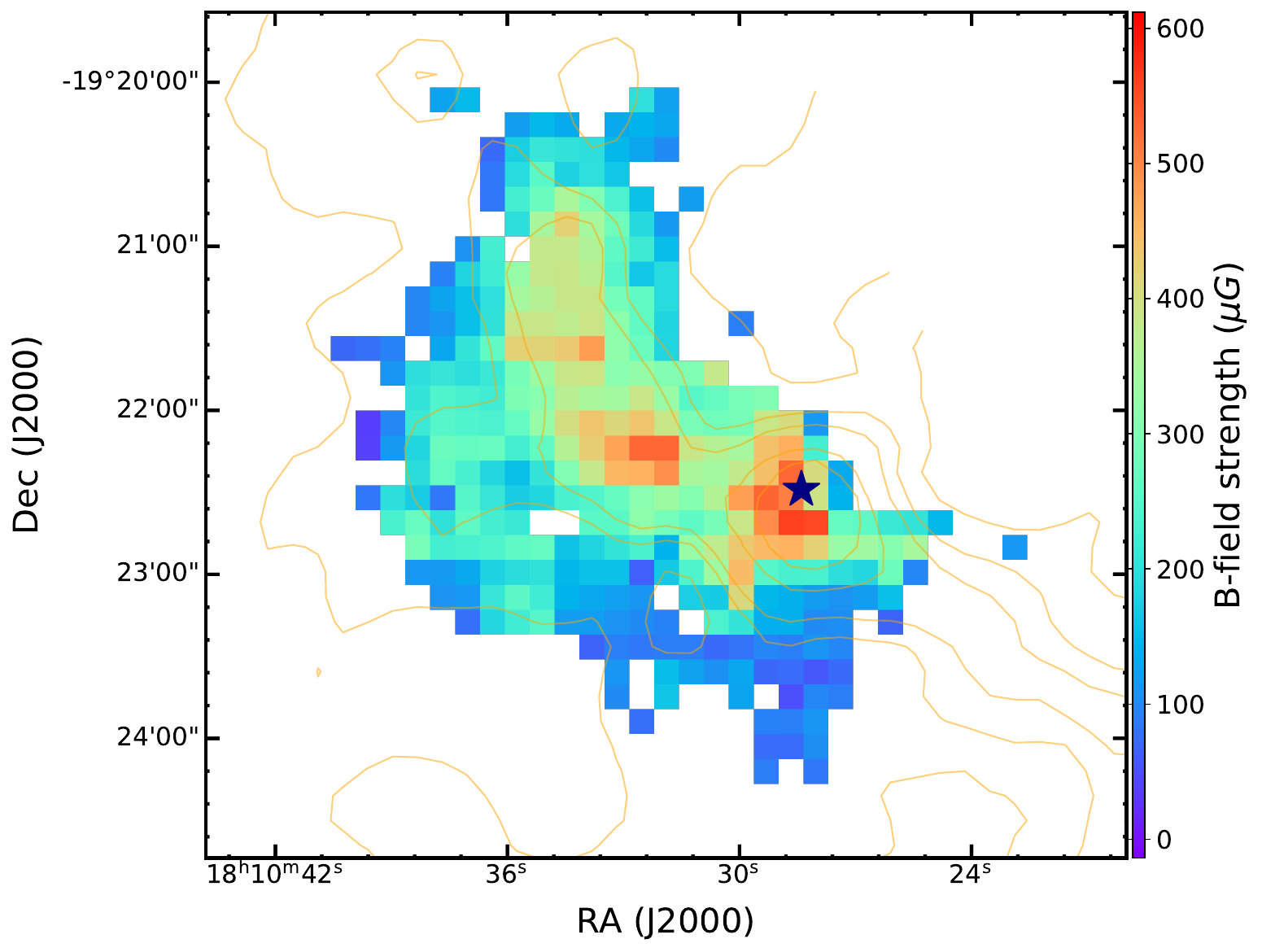}
\caption{Map of B-field strengths calculated for each pixel toward the center region of G11. The violet star indicates the location of the massive protostellar candidate (P1). The contours are the same as in Figure \ref{fig:Bfieldlength}.} \label{fig:strength} 
\end{figure}

We use Equation \ref{equ:dcf} to calculate the B-field strength for each pixel with polarization angle dispersion, $\sigma_\theta$, smaller than 25$^\circ$, as suggested by \citet{ostriker2001}. Figure \ref{fig:strength} shows the map of B-field strengths toward the center region. The strengths vary from 100 $\mu$G in the outer region to $\sim$600 $\mu$G in the filament's spine with a mean value of 235 $\mu$G.

Mapping the B-field strengths has been done only for a few objects: Orion A \citep{guerra2021orion, hwang2021jcmt}, Monoceros R2 \citep{hwang2022monr2}, and 30 Doradus \citep{tram2022dorII}. Before these studies, the mean values of B-field strengths were only estimated. Therefore, to test the robustness of the obtained B-field strengths for G11, we also use the structure function method \citep{hildebrand2009} as was done in our previous work \citep{Ngoc2021,Thuong2022} to calculate the mean B-field strength for the center region. The polarization angle dispersion obtained with the structure-function method is $\sigma_\theta=10^\circ.5\pm4^\circ.6$. The mean number density and non-thermal velocity dispersion are $n({\rm H_2})=(8.0\pm1.4) \times 10^3\;{\rm cm^{-3}}$ and $\Delta V=2.4\pm0.4\;{\rm km\;s^{-1}}$, respectively. Putting these values to Equation \ref{equ:dcf}, we obtain the mean value of $B_{\rm POS} = 242 \pm 50\;\mu{\rm G}$. This value lies well in the range of the B-field strengths calculated on the pixel-to-pixel basis which ensures the mapping process's quality.

\subsubsection{Mass-to-flux Ratios} \label{subsec:lambda}

\begin{figure}[!htb]
\centering
\includegraphics[trim=0cm 0cm 0cm 0cm,clip,width=8.5cm]{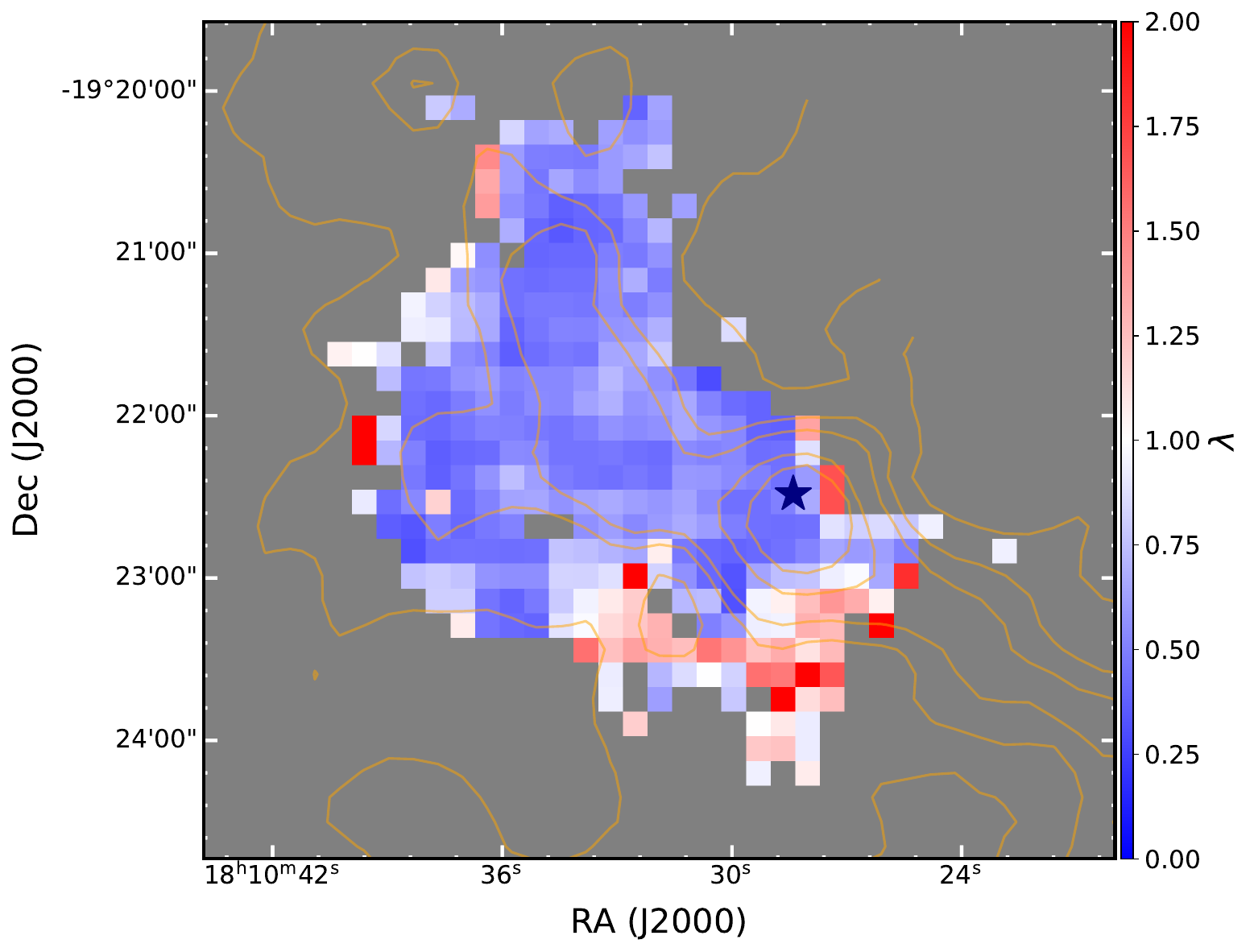}
\caption{Map of mass-to-flux-ratio toward the center region of G11. The violet star is the location of the massive protostellar candidate (P1). The contours are the same as in Figure \ref{fig:Bfieldlength}.} \label{fig:lambda} 
\end{figure}

The relative importance of gravity to B-fields is usually described by the mass-to-flux ratio, $M/\Phi$. In the units of the critical value, the mass-to-flux ratio is given by the following relation \citep{crutcher2004}:
\begin{eqnarray}\label{eq8}
\lambda=\frac{(M/\Phi)_{\rm observed}}{(M/\Phi)_{\rm critical}}=7.6\times10^{-21}\frac{N(\rm H_2)}{B_{\rm tot}}
\end{eqnarray} where $(M/\Phi)_{\rm critical}=1/(2\pi\sqrt{G})$, $G$ is the gravitational constant, $N(\rm H_2)$ is measured in cm$^{-2}$, $B_{\rm tot}$ is total B-field strengths in $\mu$G which is approximated to be $B_{\rm tot} = 1.3 \times B_{\rm POS}$ \citep{crutcher2004}.

Figure \ref{fig:lambda} shows a map of the mass-to-flux ratio toward the central region in which the colors encoded with blue represent $\lambda < 1$ and red $\lambda > 1$. It is visible that most of the region is sub-critical meaning that the gravity is not strong enough for the gravitational collapse to occur to form new stars. This is consistent with the scenario that the filament is in its early phase, accreting material from the outer region onto the filament's spine following the B-field lines. The southwestern part of the considered region is super-critical which is close to the location of the massive protostellar candidate P1 (violet star marker).

\subsubsection{Alfv\'en Mach Number} \label{subsec:Ma}

\begin{figure}[!htb]
\centering
\includegraphics[trim=4cm 1cm 4cm 2cm,clip,width=8.5cm]{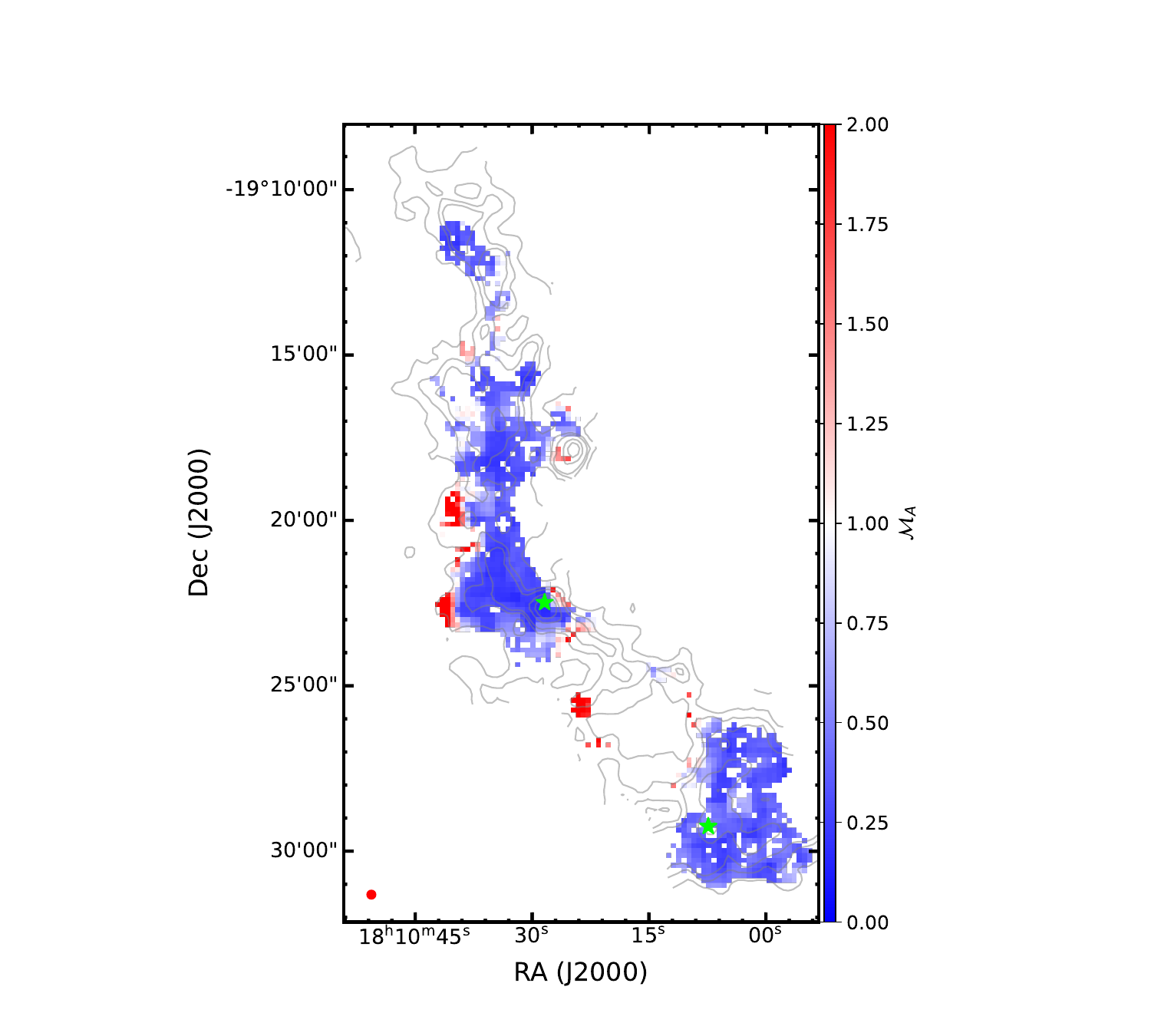}
\caption{Map of Alfv\'en Mach number. The violet stars mark the locations of the massive protostellar candidates (P1 and P6). The contours are the same as in Figure \ref{fig:Bfieldlength}.} \label{fig:ma} 
\end{figure}

The interplay between the B-fields and turbulence can be evaluated by the Alfv\'en Mach number. The Alfv\'enic Mach number, $\Ma$, of the gas is given by:

\begin{equation}\label{equa:Ma}
\Ma = \frac{\sigma_{\rm V}}{\nu_A}
\end{equation}
where $\sigma_{\rm V}$ is the one-dimensional non-thermal velocity dispersion and $\nu_A = B_{\rm tot}/\sqrt{4\pi\rho} = \mathcal{Q} \times \sigma_{\rm V}/\sigma_{\theta}$ is the Alfv\'enic velocity; both $\sigma_{\rm V}$ and $\nu_A$ are measured in km$\s^{-1}$. Combining Equation \ref{equa:Ma} and $\nu_A$, we have $\Ma = \sigma_\theta/\mathcal{Q}$. As mentioned earlier, we take the correction factor $\mathcal{Q}=0.5$ and $\sigma_\theta$ is now in radian. Equation \ref{equ:dcf} is equivalent to $\Ma \approx 3.5 \times 10^{-2} \sigma_\theta$, where $\sigma_\theta$ is measured in degrees \citep{pattle2021L1689}. 

Since the Alfv\'en Mach number depends only on the polarization angle dispersion, we produce a map of the Alfv\'en Mach number for the whole Snake filament (Figure \ref{fig:ma}). The color scale of the map is such that the blue-color represents $\Ma < 1$ and the red-color for $\Ma > 1$. The overall color of the map is blue; therefore, the region is mostly sub-Alfv\'enic. Only the outer-most regions of the filament from its spine are found to be trans- or super-Alfv\'enic. The energy of the B-fields is dominated over that of the non-thermal motions. Once again, the results are in line with the ordered B-field morphology observed in the filament. The impact of the gas turbulence is not strong enough and the B-fields are able to regulate the gas motion.

\section{Dust physics and grain properties}
\label{sec:dust}

\begin{figure*}[!htb]
\centering
\includegraphics[trim=0cm 0cm 0cm 0cm,clip,width=11.5cm]{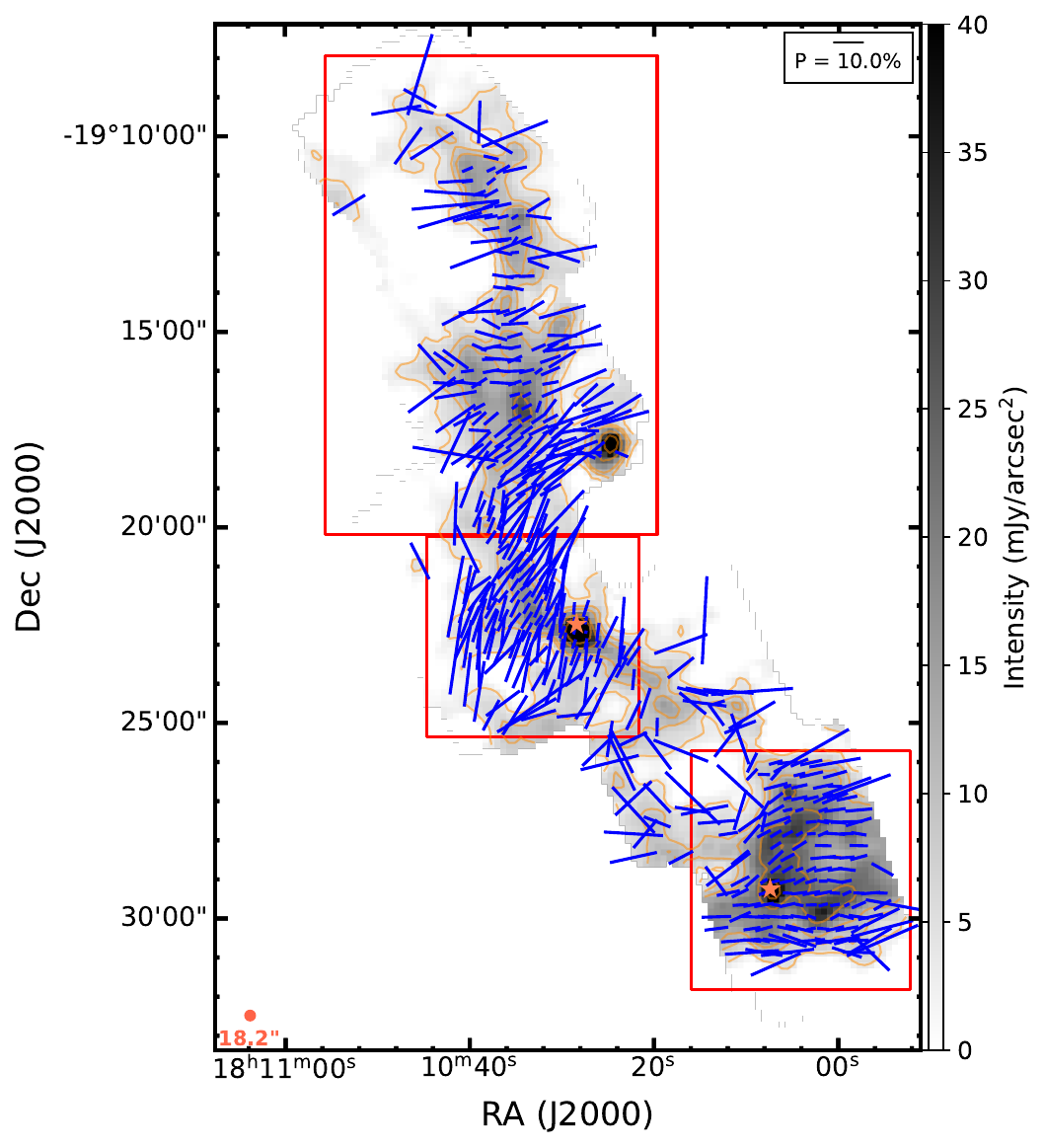}
\caption{Same as Figure \ref{fig:Bfieldlength}, but the length of polarization vectors now represents the polarization fraction. Three regions mentioned in the text are defined by the three red rectangles for the North, Center, and South regions, respectively.} \label{fig:Bfieldregions} 
\end{figure*}

In this section, we perform various analyses using polarization fraction to study grain alignment physics and dust grain properties in G11. Firstly, we analyze the relation of the obtained polarization fraction, $P$, with the intensities, $I$, gas column densities, $\NHt$, and dust temperatures, $\Td$. 

\subsection{Maps of gas densities and dust temperatures}
Gas densities and dust temperatures are the key parameters of the grain alignment mechanism based on RATs. We use the $\H_2$ column densities and dust temperatures produced by \citet{zucker2018filament} using modified blackbody fitting to the data from \textit{Herschel}'s four bands at 160, 250, 350, and 500 $\mum$.

Figure \ref{fig:nH2Td} shows the maps of gas densities (left) and dust temperatures (right). The dust temperatures decrease from $\Td>20$ K in the diffuse region to \mbox{$\Td\lesssim 15$ K} close to the filament's spine. This reveals that dust grains inside the filament are heated only by the interstellar radiation field and the contribution of the internal radiation from two protostellar-candidate sources P1 and P6 is sub-dominant.

\subsection{Polarization Fraction Map}

Figure \ref{fig:Bfieldregions} displays the B-field orientation map similar to Figure \ref{fig:Bfieldlength}, but the segment lengths are now proportional to the polarization fraction, $P$. For further studies, we divide the filament into three sub-regions, North, Center, and South, represented by the three red rectangles in Figure \ref{fig:Bfieldregions}. We note that the newly-defined Center region (Center with capital C) is slightly different from the center region encompassed by a circle shown in Figure \ref{fig:Bfieldlength} in previous sections which overlaps with the observed region of JCMT/SCUPOL. It is clearly seen that the polarization fraction is higher in the more diffuse regions far from the filament's spine but it drops significantly in the regions close to the filament's spine where the thermal emission intensities are higher. This effect, called `depolarization' or `polarization hole', is a common phenomenon that has been reported by both Optical–NIR \citep{whittet2008efficiency} and far-IR/sub-mm observations (e.g., \citealt{ward2000first}) toward molecular clouds. The origin of depolarization effect is still under debate (e.g., \citealt{pattle2019,hoang2021polhole}).

\begin{figure}[!htb]
\centering
\includegraphics[trim=0.6cm 0cm 0.6cm 0cm,clip,width=8.5cm]{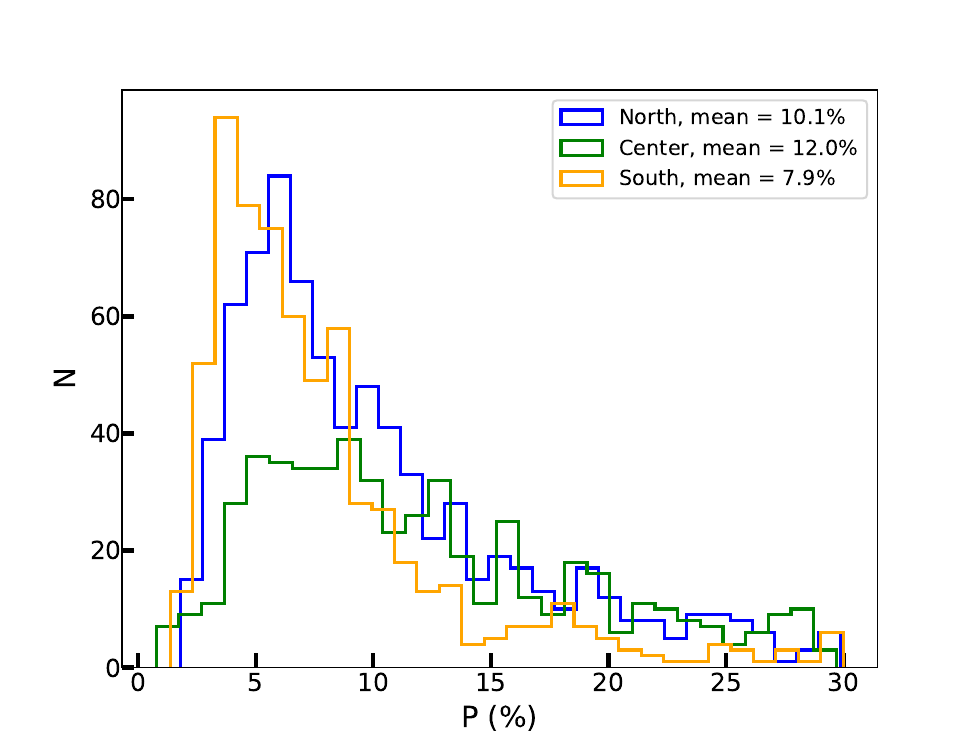}
\caption{Distributions of the polarization fractions of North (blue), Center (green), and South (orange).} \label{fig:hist_p} 
\end{figure}

Figure \ref{fig:hist_p} shows the histograms of the polarization fraction. The mean polarization fraction for the three regions is about $10\%$, but there are many pixels with high polarization above $20\%$. The Center region appears to have a higher polarization fraction than that of the North and South regions.

\subsection{Polarization Fraction versus Total Intensities} \label{subsec:PvsI}

\begin{figure}[!htb]
\centering
\includegraphics[trim=0cm 0cm 0cm 0cm,clip,width=8.5cm]{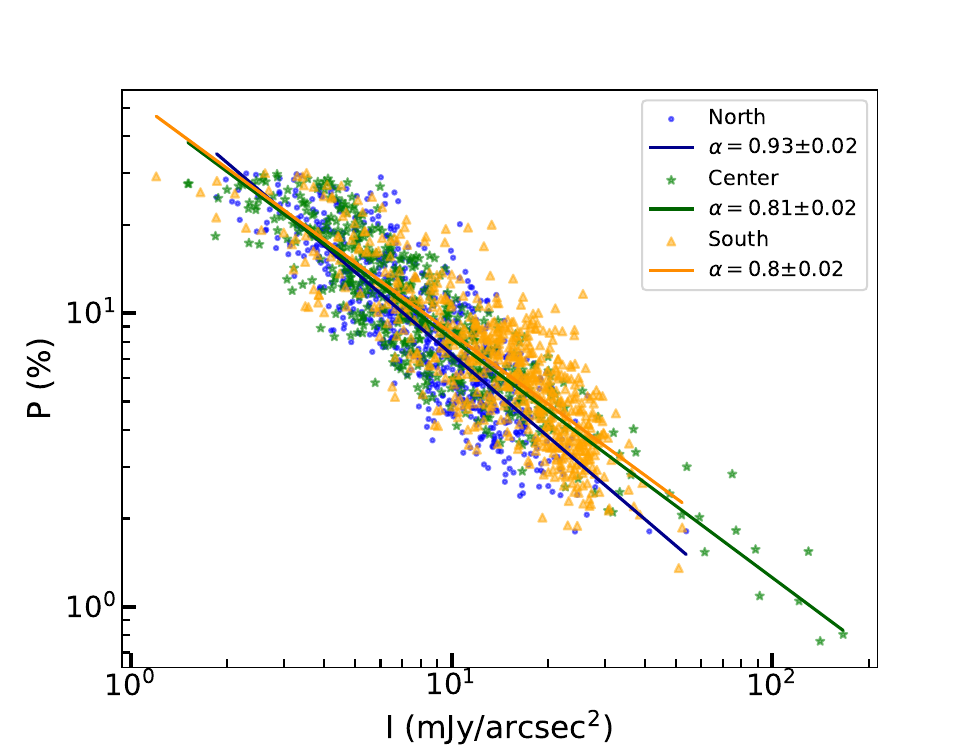}
\caption{Dependence of polarization fraction, $P(\%)$, on intensity, $I$ observed by SOFIA/HAWC+ for North (blue), center (green), and South (orange). Solid lines show the power-law fit to the data. The slope of the North is steeper than that of the Center and South.}
\label{fig:pvsI}
\end{figure}

Figure \ref{fig:pvsI} displays the variation of $P$ with $I$. The polarization fraction is very high of $\sim$10-30\% in the outer regions with intensities $I<8$ (mJy/arcsec$^2$), but it decreases with increasing total intensities toward the filament's spine. Power-law fits to this variation of the form $P \propto I^{-\alpha}$ give the power indices $\alpha = 0.93\pm0.02, 0.81\pm 0.02, 0.80\pm 0.02$ for the North, Center, and South regions, respectively. The uncertainties on $\alpha$ are only statistical and the results of the fits. We note that the value of $\alpha$ is an indication of the variation of the grain-alignment efficiency and the B-field tangling along the line of sight. In general, the expected values of $\alpha$ for molecular clouds or cores are between $1$ and $0$. The index $\alpha=1$ indicates a near-total lack of alignment in dense regions usually found for starless-core clouds, and a shallower slope with $\alpha<1$ implies that grains are still aligned in the inner regions of the cloud \citep{hoang2022internal}. In particular, the slope with $\alpha=0.5$ can be reproduced with a model of uniform grain alignment with supersonic and sub-Alfv\'enic turbulence \citep{Falceta.2008}. Turbulence is the main source of field tangling \citep{1989ApJ...346..728J,jones1992infrared,2008ApJ...679..537F}.

We found the slopes with $\alpha\sim$0.8 for the Center and South regions which are shallower than that of the North ($\alpha\sim$0.9). This means that grain alignment in the Center and South regions is more efficient than in the North. Interestingly, there exist two massive star candidates P1 and P6 in the Center and South regions, which may contribute to inducing grain alignment in the inner regions and making the slopes shallower \citep{hoang2021polhole}. The slopes of $\alpha< 1$ suggest that grain alignment is not entirely lost. We discuss the implication of found slopes and provide evidence for grain growth in the filament in Section \ref{ggrowth}.

\subsection{Polarization Fraction versus Column Densities and Dust Temperatures}
\label{subsec:PvsNH2Td}

\begin{figure*}[!htb]
\centering
\includegraphics[trim=2cm 5.5cm 2cm 5cm,clip,width=17cm]{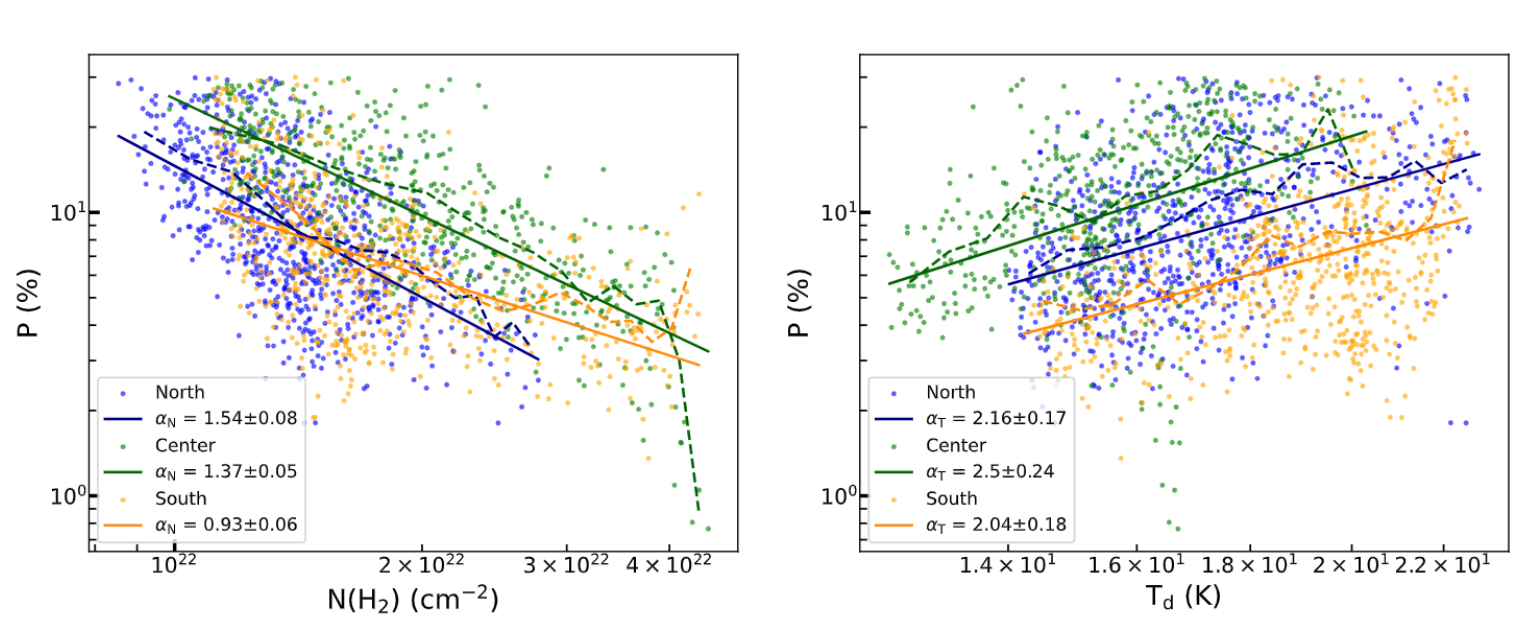}
\caption{Dependence of polarization fraction, $P(\%)$, on column densities, $\NHt$ (left) and dust temperatures, $\Td$ (right). Solid lines are the results of the fits to a power-law function (see text). The dashed lines show the mean values calculated per bin size of $1.3 \times 10^{21}$ cm$^{-2}$ for $\NHt$ and 0.6 K for $\Td$.} \label{fig:pvsNH2Td} 
\end{figure*}

The grain alignment efficiency depends on the local conditions, including the radiation field and gas densities, according to the RAT theory \citep{hoang2021polhole}. To study clearly how the local conditions of G11 affect the dust polarization fraction, in Figure \ref{fig:pvsNH2Td}, we plot the relation of $P$ with the column densities (left panel) and dust temperatures (right panel).

The $P-\NHt$ relation shows an anti-correlation which can be described by a power-law model of the form $P \propto \NHt^{-\alpha_{\rm N}}$ with slopes $\alpha_{\rm N} = 1.54 \pm 0.08$ for North, $\alpha_{\rm N} = 1.37 \pm 0.05$ for Center, and $\alpha_{\rm N} = 0.93 \pm 0.06$ for South. Similar to the $P-I$ relation, the slope $\alpha_{\rm N}$ for the North region is steeper than that for the Center and South regions, which implies the role of two massive star candidates P1 and P6 in the enhanced alignment in the latter regions.

The $P-\Td$ relation shows the correlation of the polarization fraction and the dust temperatures or equivalent to the radiation field strength. A power-law fit of the form $P \propto \Td^{\alpha_{\rm T}}$ to the observational data yields the slopes $\alpha_{\rm T} = 2.16 \pm 0.17$ for North, $\alpha_{\rm T} = 2.5 \pm 0.24$ for Center, and $\alpha_{\rm T} = 2.04 \pm 0.18$ for South. A detailed discussion on the grain alignment physics using these results is presented in Section \ref{sec:alignment}.

\subsection{Magnetic Field Tangling} \label{subsec:btangling}

\begin{figure*}[!htb]
\centering
\includegraphics[trim=1.1cm 11.2cm 1.1cm 11.2cm,clip,width=17cm]{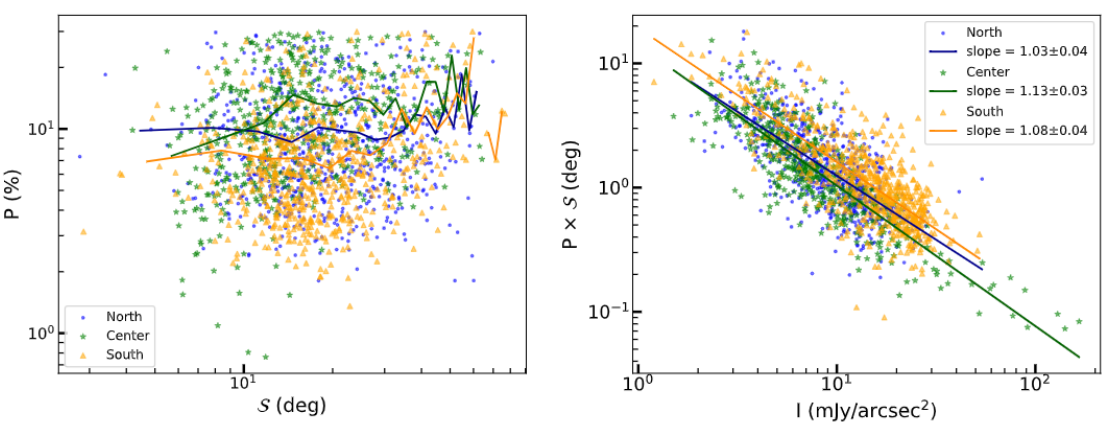}
\caption{Left: Dependence of polarization fraction, $P(\%)$, on polarization angle dispersion function, $\S$. The solid curves show the running means. Right: The variation of $P\times \S$ vs. $I$. Dashed curves show the running means and the solid lines are the results of power-law fits (see text).} \label{fig:pvsS} 
\end{figure*}

In addition to grain alignment and dust grain properties, polarization fraction, $P$, also depends on the B-field geometry along the line of sight. To disentangle the effect of B-field tangling and grain alignment on the observed dust polarization, we now analyze the polarization angle dispersion function, $\S$, and the product $P\times\S$. The product $P\times\S$ describes the averaged alignment efficiency of grains along the line of sight \citep{planck2018}. This implication is based on the observation that, for a constant grain alignment efficiency, the larger/smaller $\S$ would imply stronger/weaker B-field tangling, which produces a smaller/larger polarization fraction. Therefore, the product $P\times\S$ would provide us with information about the overall grain alignment.

We first calculate $\mathcal{S}$ using the definition described in Section 3.3 of \cite{planck2018}. For each pixel at location $x$, $\mathcal{S}(x)$ is calculated as the RMS of the polarization angle difference, $\mathcal{S}_{xi}$, of the considered pixel $x$ and pixel $i$ lying on a circle having $x$ as the center and a radius of $\delta$:

\begin{equation}\label{eq:ang_dis_func}
 \mathcal{\S}^2(x) = \frac{1}{N}
\sum _{i=1}^{N}\S_{xi}^2,
\end{equation}
where $\S_{xi}=\theta(x)-\theta(x+\delta)$ is the polarization angle difference and $N$ is the number of pixels lying on the circle. For the current study, we calculate $\S(x)$ for $\delta$ equal to two times the beam size of SOFIA/HAWC+.

Due to noise on the Stokes parameters $Q$ and $U$, $\S$ is biased. The bias of $\S$ can be positive or negative depending on whether the true value is smaller or larger than the random polarization angle $52^{\circ}$ \citep{alina2016polarization}.

The variance of angle dispersion function, $\sigma_\S$, and the debiased $\S$ is calculated as (see Section 3.5 of \citealt{planck2018}):

\begin{equation}\label{eq:sigma_s}
\begin{split}
\sigma_\S(x)& = \frac{\delta\theta(x)}{N^2\S^2} \left(\frac{1}{N}
\sum _{i=1}^{N}\theta(x+\delta)-\theta(x) \right)^2 \\
& + \frac{1}{N^2\S^2}\sum _{i=1}^{N}\delta\theta^2(x+\delta)(\theta(x+\delta)-\theta(x))^2 ,
\end{split}
\end{equation}
and
\begin{equation}
\S_{\rm db}^2(x) = \S^2(x) - \sigma_\S(x)^2, \; \; \rm {if} \;\S > \sigma_\S
\end{equation}
where $\sigma_\S(x)$ is the variance of $\S(x)$. Hereafter, we refer to $\S_{\rm db}(x)$ as $\S$ for convenience.

Figure \ref{fig:pvsS} (left) shows the relation of $P$ vs. $\S$ calculated for three different regions. The solid lines show the running means of the polarization fraction. The polarization fraction does not exhibit a clear variation with $\S$, although the data are scattered. However, in Figure \ref{fig:pvsS} (right), we show $P\times \S$ vs. $I$, which exhibits a steep decrease of $P\times\S $, i.e. the observed alignment efficiency, with increasing $I$. Thus, the B-field tangling is not the dominant cause of the depolarization in G11 but the decrease of grain alignment efficiency.

\subsection{Grain Alignment Mechanisms} \label{sec:alignment}
We now apply the modern theory of grain alignment to understand the observed dust polarization fraction and various analyses shown in the previous section.

\begin{figure}[!htb]
\centering
\includegraphics[trim=0cm 0cm 0cm 0cm,clip,width=8.5cm]{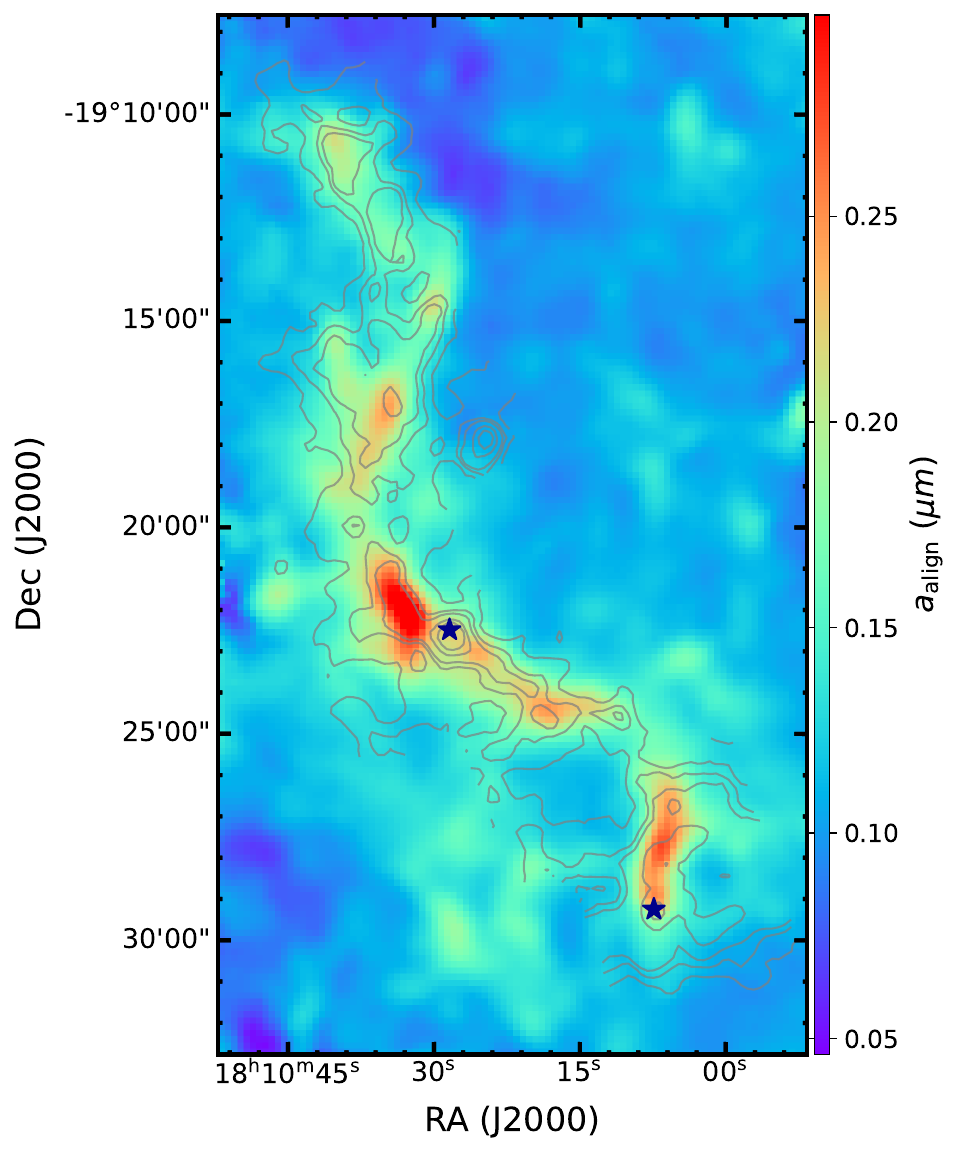}
\caption{Map of alignment sizes calculated within the framework of RATs, $a_{\rm align}$. The alignment size increases from the outer to the inner region of the filament.}\label{fig:aalign}
\end{figure}

According to the RAT theory, grains are effectively aligned only when they can rotate suprathermally, i.e., a rotation rate
much higher than the thermal value, \citep{Hoang:2008gb,Hoang.2016}. The minimum size of aligned grains, hereafter alignment size, can be calculated by the following formula \citep{hoang2021polhole},
\begin{eqnarray}
&a_{\rm align}\simeq &0.055 \hat{\rho}^{-1/7} \left(\frac{\gamma U}{0.1}\right)^{-2/7} \left(\frac{\nH}{10^{3} \cm^{-3}}\right)^{2/7} \nonumber\\
    &&\times \left(\frac{T_{\rm gas}}{10\K}\right)^{2/7} \left(\frac{\bar{\lambda}}{1.2\mum}\right)^{4/7} (1+F_{\rm IR})^{2/7},  
    \label{eq:aalign_ana}
\end{eqnarray}

where $\hat{\rho}=\rho_{\rm d}/(3 \g \cm^{-3})$ with $\rho_{\rm d}$ is the dust mass density, $\gamma$ is the anisotropy degree of the radiation field, $\bar{\lambda}$ is the mean wavelength, and $U$ is the strength of the radiation field. Above, $\nH$ is the number density of hydrogen atoms, $T_{\rm gas}$ is the gas temperature, and $F_{\rm IR}$ is the ratio of the IR damping to the collisional damping rate. As shown in Equation \ref{eq:aalign_ana}, the alignment size increases with increasing the gas density, but it decreases with increasing radiation strength, $U$ (or dust temperature).

Within the RAT paradigm, the dust polarization fraction is determined by the size distribution of aligned grains, spanning from $\aalign$ to the maximum grain size, $a_{\rm max}$ \citep{2014MNRAS.438..680H,Lee.2020}. The latter value is determined by grain growth and destruction processes. For a given $a_{\rm max}$, a larger value of $\aalign$ results in a narrower size distribution of aligned grains, which produces a lower polarization fraction, $P$. Similarly, a smaller value of $\aalign$ causes the higher $P$ because of the wider size distribution of aligned grains (see \citealt{tram2022recent}).

For calculations of alignment size in G11, we assume the local radiation field of $\gamma =0.1$, $\bar{\lambda} = 1.2 \mum$, $T_{\rm gas} = \Td$ (thermal equilibrium assumption between gas and dust is valid for dense and cold environments). The anisotropy degree of radiation $\gamma =0.1$ is typical for the diffuse interstellar radiation field \citep{1997ApJ...480..633D,2007ApJ...663.1055B}. G11 has no nearby bright star, the dust is mainly heated by the interstellar radiation field, therefore, $\gamma =0.1$ is chosen. We use $\Td$ and $\nH = 2n(\rm{H_2})$ maps of G11 shown in Figure \ref{fig:nH2Td}. To obtain $U$, we use the relationship between the dust temperature and the radiation strength for silicate grains having sizes in the range of 0.01 to 1 $\mu$m with dust heating and cooling balance, and radiation strength $U<10^4$ ($\sim 75$ K): $U \approx (\Td/{16.4\;\rm K})^6$ \citep{draine2010}.

Figure \ref{fig:aalign} shows the map of the alignment sizes. The alignment sizes increase from the outer layer to the filament's spine as the consequence of the increase of the gas density and the decrease of the dust temperature (see Figure \ref{fig:nH2Td}). Compared to the map of the polarization fraction (see Figure \ref{fig:Bfieldregions}), one can see the overall anti-correlation of $a_{\rm align}$ and $P$ as expected.

\begin{figure}[!htb]
\includegraphics[trim=0cm 0cm 0cm 0cm,clip,width=8.5cm]{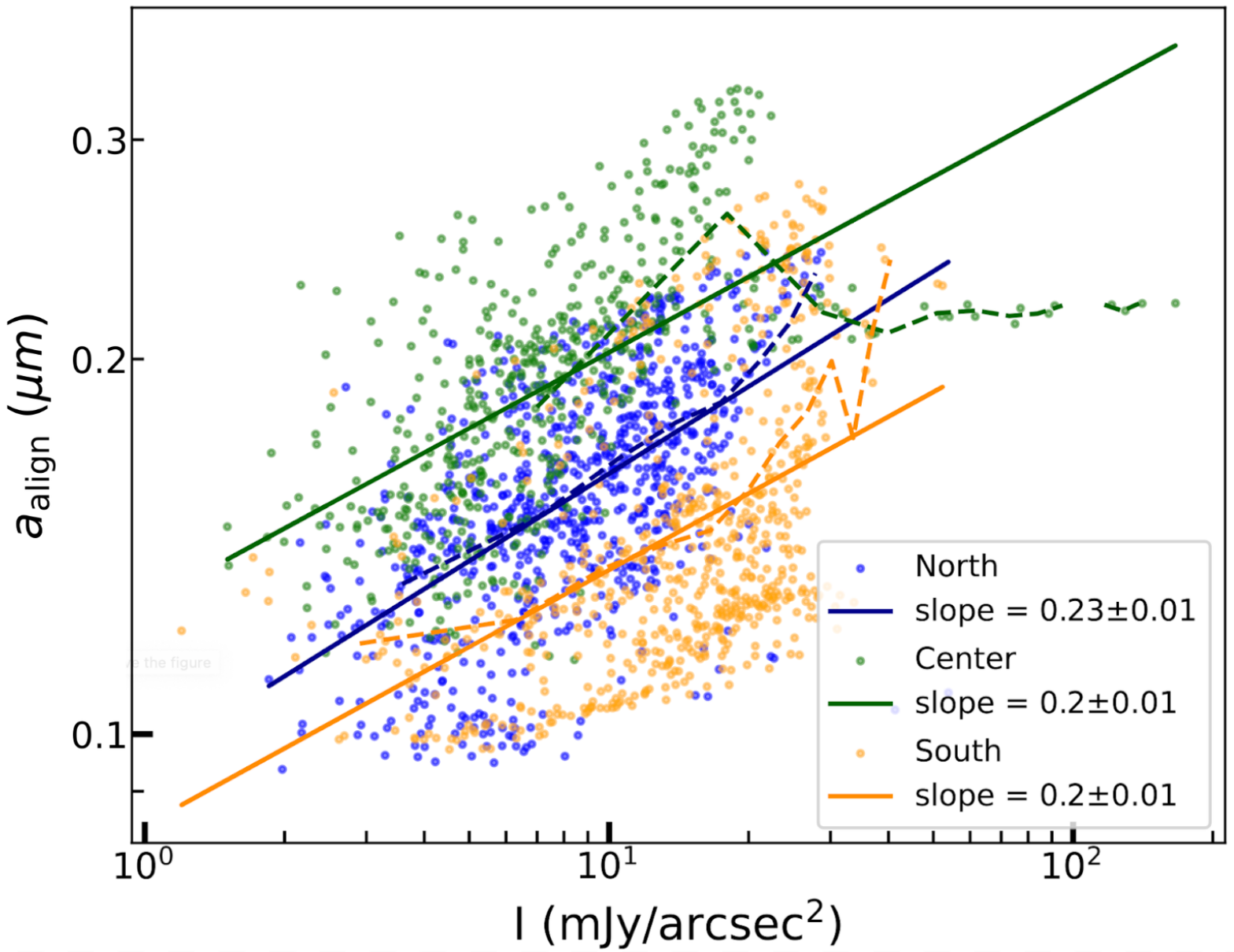}
\caption{Variation of the alignment size, $a_{\rm align}$, with intensity, $I$. The dashed lines show the running mean values. Solid curves are the power-law fits to the data for the three regions. }\label{fig:alignvsI} 
\end{figure}

Upon closer inspection of the variation of the polarization fraction on the alignment size, we plot the variation of $\aalign$ with $I$ in Figure \ref{fig:alignvsI}. The alignment size increases with the intensity with slightly shallower slopes for the Center and South regions.

\begin{figure*}[!htb]
\includegraphics[trim=0cm 5.2cm 0.6cm 5.2cm,clip,width=17cm]{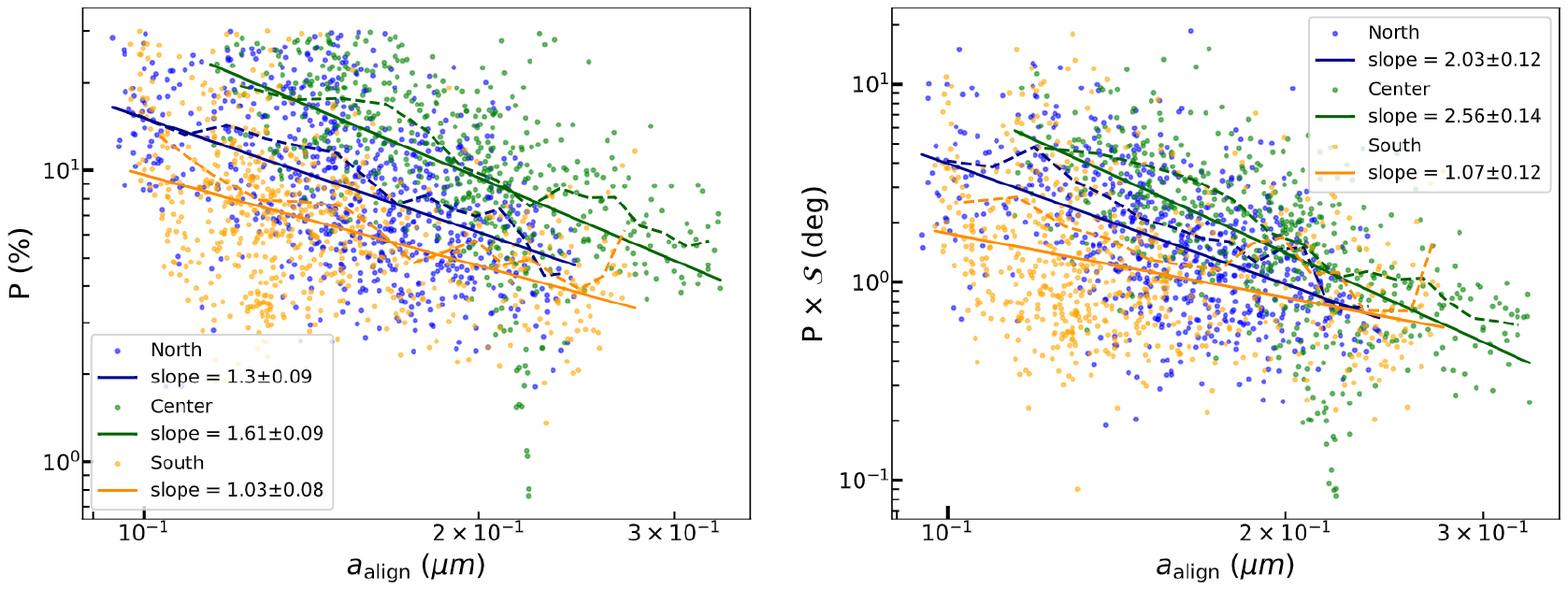}
\caption{The variation of the polarization fraction (left) and of the alignment efficiency, $P\times\S$, (right) on the alignment size, $a_{\rm align}$. The dashed lines show the running means and solid lines are the results of the fits to a power-law function (see text).} \label{fig:Pvsalign} 
\end{figure*}

Figure \ref{fig:Pvsalign} shows the variation of $P$ (left) and $P \times \S$ (right) with $a_{\rm align}$. The decrease in polarization fraction and alignment efficiency is visible with increasing the alignment size by RATs. 
From numerical modeling, it is found that, as $a_{\rm align}$ increases, the polarization fraction decreases due to the reduction of the fraction of aligned grains \citep{Lee.2020,hoang2021polhole}. Hence, the depolarization in G11 could be explained by the decrease in the RAT alignment efficiency toward the high column density with low dust temperature regions.

\subsection{Magnetic Relaxation}

\begin{figure}[!htb]
\centering
\includegraphics[trim=0cm 0cm 0cm 0cm,clip,width=8.5cm]{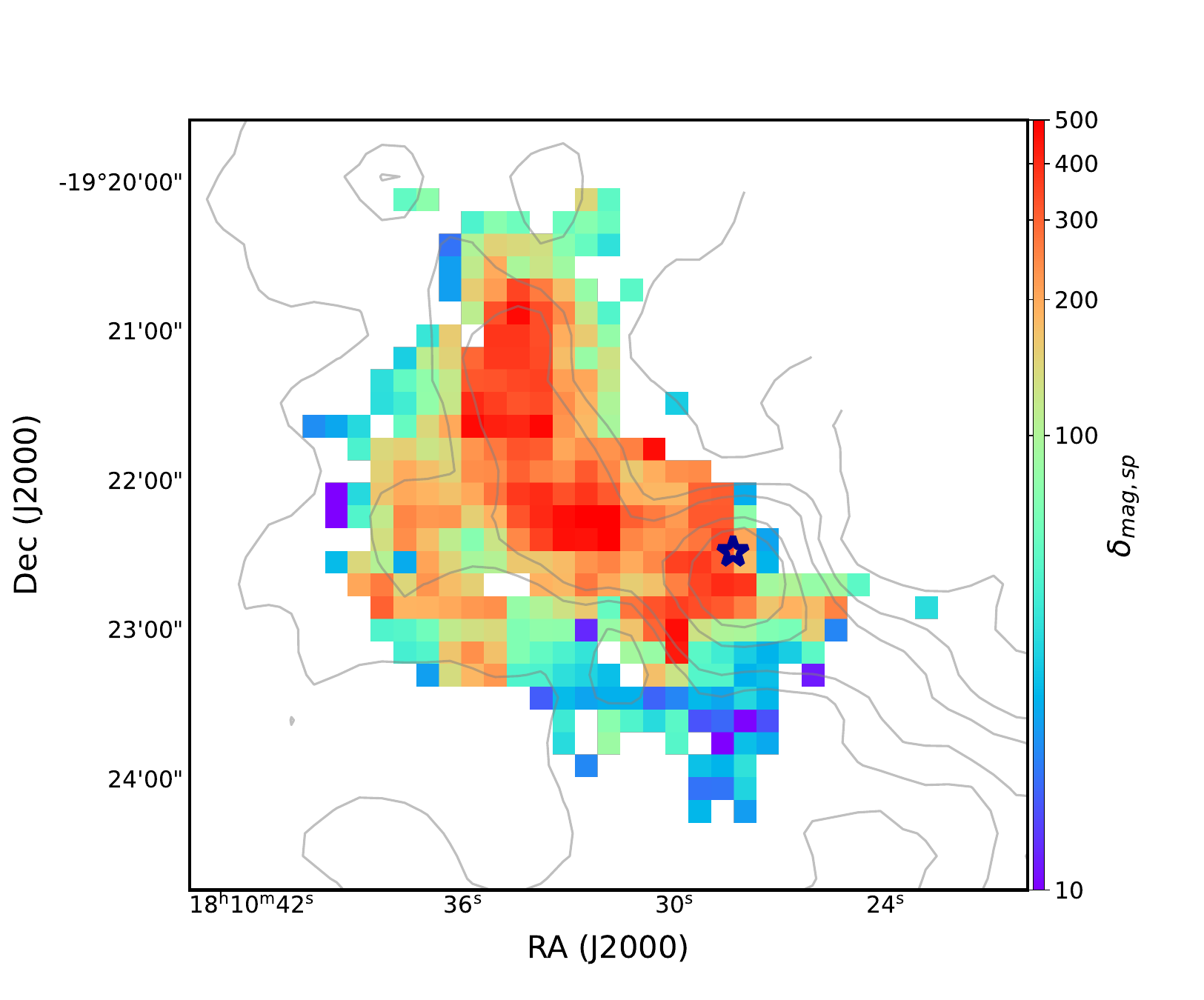}
\caption{Map of $\delta_{\rm mag,sp}$ for the center region of G11. The value of $\delta_{\rm mag, sp}$ increases from the outer to the inner region and is larger than $\sim10$. The star marker shows the location of P1.} \label{fig:relaxation}
\end{figure}

The magnetic properties of dust are essential for their interaction and alignment with the ambient B-fields. Dust grains containing embedded iron atoms make interstellar dust a natural paramagnetic material. Dust grains become superparamagnetic when iron atoms are distributed as clusters inside the grains (e.g., \citealt{hoang2022internal}). The alignment of the grain angular momentum with the B-fields is usually thought due to paramagnetic relaxation \citep{davis1951polarization}. Nevertheless, the paramagnetic relaxation alone cannot produce efficient alignment and is insufficient to explain the observational data \citep{Hoang.2016}. In the presence of RATs, magnetic relaxation, general case of relaxation due to magnetic dissipation for any type of magnetic material (para, super-para, ferro-magnetic, etc.), can enhance the alignment degree, known as magnetically enhanced RAT (or MRAT) alignment mechanism \citep{Hoang.2016}.

The strength of magnetic relaxation, $\delta_{\rm mag}$, is defined by the ratio of the gas collision damping timescale, $\tau_{\rm gas}$ to magnetic relaxation time, $\tau_{\rm mag,sp}$. For grains with embedded iron clusters which are plausible for grains in dense regions due to grain evolution, one has
\begin{equation}
    \delta_{\rm mag,sp} = \frac{\tau_{\rm gas}}{\tau_{\rm mag,sp}} =56a_{-5}^{-1}\frac{N_{cl}\phi_{\rm sp,-2}\hat{p}^2B_3^2}{\hat{\rho}n_4T_{\rm gas,1}^{1/2}}\frac{k_{\rm sp}(\Omega)}{T_{\rm d,1}},\label{eq:delta_m}
\end{equation}
where $a_{-5}=a/(10^{-5}\;{\rm cm})$, $B_{3}=B_{\rm tot}/(10^{3}\;\mu {\rm G})$, $n_{4}=n_{\rm H}/(10^{4}\;{\rm cm^{-3}})$ with $n_{\rm H}\approx 2n(\rm {H_{2}})$ for molecular gas, $T_{\rm gas,1}=T_{\rm gas}/(10\K), T_{{\rm d,1}}=T_{\rm d}/(10\K)$, $\hat{p}=p/5.5$ with $p\approx 5.5$ the coefficient describing the magnetic moment of an iron atom, $N_{\rm cl}$ is the number of iron atoms per cluster, $\phi_{\rm sp}$ is the volume filling factor of iron clusters with $\phi_{\rm sp,-2}=\phi_{\rm sp}/0.01$, and $k_{\rm sp}(\Omega)$ is the function of the grain rotation frequency $\Omega$ which is of order unity (see \citealt{hoang2022internal} for details). 

Magnetic relaxation is considered effective when $\delta_{\rm mag,sp}>1$, i.e., relaxation occurs faster than the gas rotational damping. Numerical calculations in \cite{Hoang.2016} show that the joint action of suprathermal rotation by RATs and magnetic relaxation can enhance the alignment degree so that superparamagnetic grains can achieve perfect alignment for $\delta_{\rm mag,sp}>10$.

To study whether magnetic relaxation has any effects on RAT alignment in this filament, we use the maps of the gas densities, $n_{\rm H}$, B-field strengths, $B_{\rm tot}$, and dust temperature, $\Td$ and plug their values into Equation (\ref{eq:delta_m}) to obtain $\delta_{\rm mag,sp}$ for the entire filament. We assume the typical value of $N_{\rm cl} = 100$ and $\phi_{\rm sp}=0.01$ (i.e., about $3\%$ of iron abundance in the form of iron clusters, see \citealt{Hoang.2016}). Figure \ref{fig:relaxation} shows the results. As shown, even for the chosen parameters with a low level of iron inclusions, one has $\delta_{\rm mag,sp}\gtrsim 10$ in the entire filament, indicating that magnetic relaxation is very efficient in the filament so that grains can achieve perfect alignment due to the MRAT mechanism (\citealt{hoang2022internal}). More discussions on magnetic relaxation in the region are presented in Section \ref{mrelaxation}.

Finally, we note that, although $\delta_{mag,sp}$ increases toward the inner region of the filament (see Figure \ref{fig:relaxation}), the expected polarization degree does not necessarily increase because of the saturation of MRAT alignment efficiency and the increase in the minimum alignment size $a_{\rm align}$ (see the previous subsection). 

\section{Discussions} \label{sec:dis}
\subsection{Characteristics of B-fields and Comparison with Previous Studies}
B-fields are thought to play an important role in the formation and evolution of interstellar filament and regulate star formation \citep{mckee2003formation,henney2009radiation}. Previous studies have revealed that B-fields tend to be parallel to the filament in the diffuse regions of low column densities, and they become perpendicular to the filament in dense regions of high column density with $N_{\rm H}\gtrsim 3$-$5\times 10^{21}\cm^{-2}$ \citep{Planck.2016,Soler.2017,Soler.2017vv}.

In this paper, we derived the B-field orientation map by using polarimetric data taken by SOFIA/HAWC+. We found that the B-fields in G11 are mostly perpendicular to the filament's spine, in particular at the highest column densities (see Figure \ref{fig:orienNH2}). From Figure \ref{fig:orienNH2}, one can see that our observations mainly trace the dense region with $\NHt> 10^{22}\cm^{-2}$. Therefore, our finding of the perpendicular orientation between B-fields and the filament is in agreement with previous studies \citep{Planck.2016,Soler.2017,Soler.2017vv}.

We applied the DCF method to calculate the map of the B-field strengths of the center region. The B-field strengths vary from $100$-$600\;\mu$G and are strongest close to the filament's spine at high densities. The B-field strengths decrease when going from the filament's spine to the outer regions. A previous study by \citet{pillai2015magnetic} using JCMT/SCUPOL measured the lower limit of the B-field strength toward the densest region of the center region (see Figure \ref{fig:bfieldscupol}) to be $\sim$$200\;\mu$G. This value is in the range of our measured B-field strengths and in an excellent agreement with our average value $B_{\rm POS} = 242 \pm 50\;\mu{\rm G}$ estimated from the structure function method.

The B-field strengths found for G11 are also comparable to the measurements in other massive filaments, for example, \mbox{$\sim$$50$ $\mu$G} for G$35.39-0.33$ \citep{liu2018G35}, \mbox{$\sim$$100$-$800$ $\mu$G} for NGC 6334 \citep{arzoumanian2021NGC6334}, \mbox{$\sim$$60$-$470$ $\mu$G} for G34.43+0.24 \citep{soam2019G34}, \mbox{$\sim$$20$-$100$ $\mu$G} for G47.06+0.26 \citep{stephens2022G47}, and \mbox{$\sim$$600$-$1000$ $\mu$G} for DR 21 \citep{Ching.20220u}.

We quantified the relative role of B-fields compared to gravity and turbulence by calculating the mass-to-flux ratio, $\lambda$, and Alfv\'enic Mach number, $\Ma$. We found that the central region is sub-critical, implying that gravity is not strong enough to cause gravitational collapse. Moreover, we found $\Ma<1$ over most parts of the filament, which implies that B-fields dominate over turbulence. This finding is consistent with the idea that the filament is still in its early stages and the influence of B-fields on its evolution is important.

\subsection{Polarization Hole and RAT Alignment}
Observations of dust polarization toward different environments usually report the decrease of polarization fraction with increasing total intensity, which is usually referred to as polarization hole (see, e.g., \citealt{pattle2019} for a review). A detailed discussion on the origin of the polarization hole using RAT grain alignment is presented in \cite{hoang2021polhole}.

G11 without bright embedded sources is an ideal target to provide evidence for the RAT alignment mechanism because the main radiation source for grain heating and alignment is from interstellar radiation fields \citep{hoang2021polhole}. We found the slope of $P\propto I^{-\alpha}$ with $\alpha\sim$$0.8$-$0.9$ for G11, which is much steeper than observed in the massive filament DR21 by JCMT/POL2 with $\alpha\sim$$0.3$ \citep{Ching.20220u} and Serpens South by SOFIA/HAWC+ with $\alpha\sim$$0.5$ \citep{2020NatAs.tmp..159P}. The steep slope in G11 reveals that grain alignment is only efficient in the outer region and becomes significantly lost in the inner region.

In Section \ref{subsec:btangling}, we prove that the depolarization by B-field tangling in this dense filament appears to be a minor effect and it must be due to grain alignment. To understand the role of grain alignment in causing the polarization-hole effect, we calculated the minimum size for grain alignment by RATs, $a_{\rm align}$, as a function of the local gas density and radiation field (dust temperature) using the RAT theory \citep{hoang2021polhole}. We found a tight correlation between the alignment size with the total intensity (Figure \ref{fig:alignvsI}). Moreover, we found a tight anti-correlation of both polarization fraction, $P$, and the observed alignment efficiency, $P\times\S$, with $a_{\rm align}$ (Figure \ref{fig:Pvsalign}), implying the reduced range of alignment size toward the filament's spine be the potential cause of the polarization hole, as expected in the RAT theory \citep{hoang2021polhole}. Indeed, dust grains having sizes above $a_{\rm align}$ must have their Larmor precession be faster than the gas damping timescale in order to be stably aligned with B-fields \citep{tram2022recent}. We checked this condition for G11 by calculating the maximum grain alignment sizes for dust grains in the densest regions of G11's spine using Equation 25 of \cite{tram2022recent}. We found that even paramagnetic grains of sizes up to $\sim$100$\mum$ can still be efficiently aligned with B-fields. Therefore, Larmor precession is not the reason for the depolarization but the reduced RAT alignment efficiency toward the denser region, in contrast to the case of protostellar cores \citep{hoang2022internal,Giang.2022}.

\subsection{On the Role of Magnetic Relaxation on RAT Alignment}\label{mrelaxation}
Our observational data reveal that the polarization fraction is very high in the outer region of the filament, spanning between $20$-$30\%$ (see Figures \ref{fig:hist_p} and \ref{fig:pvsI}). This polarization level is much higher than the average level of the ISM of $P\sim$$15\%$ observed by \citet{planck2018}, but it is comparable to the observations toward the massive filament DR21 \citep{Ching.20220u} and G34.43 \citep{soam2019G34}. Previous modeling of Planck data concluded that a high alignment degree of grains is required to reproduce the observational data \citep{Guillet.2018,Hensley.2023}. Therefore, we expect that the higher polarization level observed in the outer regions of massive filaments can be achieved only if grains can be perfectly aligned. 

The perfect alignment of grains cannot be achieved by RATs only due to the dependence of the alignment efficiency on the different parameters such as the angle between the radiation direction and the B-field, the grain shape, and composition \citep{Hoang.2016,Herranen.2021}. However, the effect of enhanced magnetic relaxation by grains with iron inclusions is predicted to increase the RAT alignment efficiency \citep{Hoang.2016,hoang2022internal}. Using the map of B-field strengths obtained by the DCF method, for the first time, we observationally estimate the importance of magnetic relaxation and MRAT alignment mechanism in a filament. Due to strong B-fields in G11, we found that a small level of iron inclusions (i.e., $N_{\rm cl}=100$ and $\phi_{\rm sp}=0.01$) can produce the magnetic relaxation faster than the gas randomization with the magnetic relaxation parameter $\delta_{\rm mag,sp}>10$ (see Figure \ref{fig:relaxation}). In dense regions like the inner filaments close to their spines, one expects the incorporation of iron inclusions into dust grains due to grain collisions and thus larger $\delta_{\rm mag,sp}$. The significantly enhanced magnetic relaxation combined with RATs can induce perfect alignment for grains with iron inclusions, which could reproduce the high polarization fraction observed toward G11. The importance of MRAT is also reported in Galactic Center by \cite{Akshaya.2023pmi}.

\subsection{Implications for Grain Growth} \label{ggrowth}
Observational studies suggest that the upper limit of the grain size distribution in the ISM is $\sim$$0.25\mum$ \citep{Mathis.1977} (so-called MRN distribution). However, grain growth is expected to occur in dense molecular clouds. Dust polarization is a useful tracer for grain growth. Previous observations of starlight polarization combined with numerical modeling using the RAT theory in \cite{Vaillancourt:2020ch} reveal grain growth in dense clouds.

In this paper, based on the map of alignment size (Figure \ref{fig:aalign}) and the polarization fraction (Figure \ref{fig:Bfieldregions}), we constrain the lower limit for the maximum grain size, $a_{\rm max}$, of the grain size distribution that is required to reproduce the polarization fraction slope. The outer region has a high polarization fraction (Figure \ref{fig:Bfieldregions}), therefore, $a_{\rm max}$ has to be much larger than $\aalign \sim$$0.05$-$0.15\;\mum$ ( typical values of $\aalign$ in the outer region, see Figure \ref{fig:aalign}). This can be satisfied with the ISM value of $a_{\rm max}\sim$$0.25\mum$ from the MRN distribution. In particular, the estimated slope of $\alpha\sim$$0.8$-$0.9< 1$  (Figure \ref{fig:pvsI}) implies that grain alignment is not completely lost ($\alpha=1$), even in the filament's spine where the desities are highest. Therefore, there still exist grains that can emit polarized radiation. This is only satisfied when grain growth could occur within the filament, which increases the maximum grain size beyond the $\aalign$ values of the filament's spine of 0.30 $\mu$m, namely $a_{\rm max}>\aalign\sim$0.30 $\mu$m (Figure \ref{fig:aalign}). Thus, we found that grain growth already occurs in the G11 filament with moderate densities, $n{\rm (\H_2)}$, of the order of $\sim$$10^{4}\cm^{-3}$ (Figure \ref{fig:nH2Td}, left panel).

\section{Conclusions} \label{sec:conclusions}
In this paper, we report a comprehensive analysis of the polarization data toward G11 taken by the polarimeter SOFIA/HAWC+ at 214 $\mum$ wavelength. We study the B-fields'properties and grain alignment's physics in this region. Our main results are summarized as follows:
\begin{enumerate}
\item 
We constructed the map of B-fields by rotating the dust polarization angles by $90^\circ$. The B-fields are mainly perpendicular to the filament's spine. There is a slow turning of the relative angles between the B-fields and the filament's spine from small angles in the lower density regions to perpendicular in the high-density regions (Figure \ref{fig:orienNH2}).
\item 
We derived the map of B-field strengths for the center region of the filament using the DCF method. The B-field strengths vary from \mbox{100-600 $\mu$G} with the highest strengths close to the filament's spine of high density and lower strengths in the outer regions of the filament of lower density.
\item 
We calculated the mass-to-flux ratio and Alfv\'enic Mach number maps in the center region. These maps show that G11 has strong B-fields, which are dominant over turbulence. The central region of the filament is mostly sub-critical. 
\item 
We performed different analyses using dust polarization fraction for the entire Snake filament to constrain grain alignment physics and dust properties. Using the RAdiative Torque Alignment theory, we found that the increase of minimum sizes of dust grains toward the filament center by RAT successfully explains the decrease of the polarization fraction with increasing gas column densities and total intensities. From the alignment efficiency studies taking into account the contribution of the polarization angle dispersion function, the B-field tangling seems to be minuscule and cannot explain the depolarization effect in the filament's spine.
\item 
We constrained the grain growth using the slopes of the polarization fraction vs. intensities and RAT alignment theory. In the outer regions, the maximum grain sizes, $a_{\rm max}$, are larger than $\aalign \sim$$0.1 \mum$. In the filament's spine, $a_{\rm max}$ must be larger than the alignment sizes $\aalign\sim$0.30$\mum$ to reproduce the slope of $\alpha\sim$0.8-0.9.
\item 
Using the B-field strengths measured by the DCF method, we evaluated the importance of magnetic relaxation on RAT alignment. We found that grains can be perfectly aligned by the joint effect of enhanced magnetic relaxation and RATs, which can successfully explain the high polarization fraction in the outer region of G11 with $P\gtrsim 20\%$.  
\end{enumerate}

We are grateful to the anonymous reviewer for her/his detailed comments that helped improve the clarity of the paper. This research is based on observations made with the NASA/DLR Stratospheric Observatory for Infrared Astronomy (SOFIA). SOFIA is jointly operated by the Universities Space Research Association, Inc. (USRA), under NASA contract NNA17BF53C, and the Deutsches SOFIA Institut (DSI) under DLR contract 50 OK 0901 to the University of Stuttgart. The James Clerk Maxwell Telescope is operated by the East Asian Observatory on behalf of The National Astronomical Observatory of Japan, Academia Sinica Institute of Astronomy and Astrophysics in Taiwan, the Korea Astronomy and Space Science Institute, the National Astronomical Observatories of China, and the Chinese Academy of Sciences (grant No.XDB09000000), with additional funding support from the Science and Technology Facilities Council of the United Kingdom and participating universities in the United Kingdom and Canada. N.B.N. and P.N.D. are funded by Vietnam National Foundation for Science and Technology Development (NAFOSTED) under grant number 103.99-2019.368 and Vietnam Academy of Science and Technology (VAST) under grant number NCVCC39.02/22-23. T.H. acknowledges the support from the National Research Foundation of Korea (NRF) grant funded by the Korea government (MSIT) (2019R1A2C1087045). This work was partly supported by a grant from the Simons Foundation to IFIRSE, ICISE (916424, N.H.). N.B.N. was funded by the Master, Ph.D. Scholarship Programme of Vingroup Innovation Foundation (VINIF), code VINIF.2022.TS083. DDN is grateful to the LABEX Lyon Institute of Origins (ANR-10-LABX-0066) Lyon for its financial support within the program ``Investissements d'Avenir'' of the French government operated by the National Research Agency (ANR). N.L. acknowledges support from the First TEAM grant of the Foundation for Polish Science No. POIR.04.04.00-00-5D21/18-00 (PI: A. Karska). We would like to thank the ICISE staff for their enthusiastic support. 
\software{Starlink \citep{currie2014}, Astropy \citep{robitaille2013astropy}, Aplpy \citep{robitaille2012aplpy,robitaille2019aplpy}}
\facilities{Stratospheric Observatory for Infrared Astronomy (SOFIA), James Clerk Maxwell Telescope (JCMT), Herschel Space Observatory}

\bibliography{snake}{}

\bibliographystyle{aasjournal}

\end{document}